\documentclass[aps,prd,preprint,superscriptaddress,nofootinbib,amsmath,amssymb]{revtex4}
\usepackage{graphicx}
\usepackage{dcolumn}
\usepackage{bm}
\usepackage{amsmath}
\usepackage{amsfonts}
\usepackage{amssymb}
\usepackage{appendix}
\usepackage{pdfpages}
\usepackage{graphicx}
\usepackage{wrapfig}
\usepackage{multirow}
\usepackage{mathrsfs}
\usepackage{epstopdf}
\usepackage{slashed}
\usepackage{soul}
\usepackage{mathtools}
\usepackage{floatrow}
\usepackage{float}
\usepackage{csquotes}
\usepackage{wrapfig}
\usepackage{graphicx}
\usepackage{makecell}
\usepackage{diagbox}
\usepackage{makecell}
\usepackage{multirow}
\usepackage{pifont}
\usepackage{amssymb}
\usepackage{epsfig}
\usepackage[utf8]{inputenc}
\usepackage{epsfig}
\usepackage{dcolumn}
\usepackage{morefloats}
\usepackage{url}
\usepackage{hyperref}
\hypersetup{
  colorlinks=true,
  citecolor=green,
  linkcolor=blue,
  urlcolor=blue}
\newcommand{\xmark}{\ding{53}}
\newcommand{\ba}{\begin{array}}
\newcommand{\ea}{\end{array}}
\def\be{\begin{equation}}
\def\ee{\end{equation}}
\def\bea{\begin{eqnarray}}
\def\eea{\end{eqnarray}}
\def\gsim{\ \rlap{\raise 2pt\hbox{$>$}}{\lower 2pt \hbox{$\sim$}}\ }
\def\lsim{\ \rlap{\raise 2pt\hbox{$<$}}{\lower 2pt \hbox{$\sim$}}\ }
\def\dslash{\kern-4pt \not{\hbox{\kern-2pt $\partial$}}}
\def\pslash{\not{\hbox{\kern-2pt p}}}

\begin{document}
\DeclareGraphicsExtensions{.eps,.ps}


\title{\boldmath \color{blue} Consequences of  $ \mu-\tau $ Reflection Symmetry at DUNE}



\author{Newton Nath}
\email[Email Address: ]{newton@ihep.ac.cn}
\affiliation{
Institute of High Energy Physics, Chinese Academy of Sciences, Beijing, 100049, China}
\affiliation{
School of Physical Sciences, University of Chinese Academy of Sciences, Beijing, 100049, China}
\begin{abstract}
{\noindent
We  consider minimal type-I seesaw framework  to realize $ \mu-\tau $ reflection symmetry in the low-energy neutrino mass matrix, $ M_{\nu} $.
 Considering  DUNE experiment, we scrutinize its potential to measure the precision of 2-3 mixing angle, $ \theta_{23} $ and the Dirac CP-phase, $ \delta $ for the given symmetry. Later, we examine the precision of these two parameters considering NuFit-3.2
  data as one of the important true points. To study the low-energy phenomenology, we further discuss various breaking patterns of such an exact symmetry. Moreover,  for each breaking scenario we perform  the capability test  of DUNE for the determination of  $ \theta_{23} $ and to establish the phenomenon of CP violation considering the true benchmark point arising from the breaking of $ \mu-\tau $ reflection symmetry. We also make remarks on the potential of DUNE to rule out maximal CP-violation or CP-conservation  hypothesis  at a certain confidence level for different scenarios.
}
\end{abstract}
\maketitle

\section{Introduction}
During last a few years, there has been remarkable progress  in the field of neutrino physics which guided us to understand some intriguing aspects of neutrinos in a comprehensive manner. It is now a well established phenomenon from different  experimental results that neutrinos possess non-zero mass, and their  different flavors are mixed \cite{Patrignani:2016xqp}.
However, the dynamical origin associated with neutrinos mass generation as well as mixing patterns are still unknown. There have been numerous theoretical attempts to understand the nature of tiny neutrino masses, among which the \textit{seesaw mechanism} is considered to be the  highly appreciated one \cite{Minkowski:1977sc,Yanagida:1979as,GellMann:1980vs,Mohapatra:1979ia,Schechter:1980gr}.
The simplest way to generate neutrino masses is to add at least two $ SU(2) $ singlet right-handed neutrino fields (i.e. $N_{\mu R}, N_{\tau R} $ ) in the Standard Model (SM). The relevant SM gauge invariant Lagrangian containing the neutrino Yukawa matrix and the Majorana neutrino mass matrix can be written as
\begin{equation}\label{eq:lag}
-\mathcal{L}\supset  \overline{L}_{\alpha L}~Y_\nu^{} N_R \widetilde{H} + \dfrac{1}{2}\overline{N^C_R}M_R N_R + \mathrm{h.c.} \; ,
\end{equation}
where $L_{\alpha L}^{} = (\nu_{\alpha},~ \alpha)^{T}_L$ is the left-handed lepton doublet, $Y_\nu^{}$ denotes the neutrino Yukawa matrix, $\widetilde{H} = i \sigma_2^{} H^*$ with $H$ being the Higgs doublet in the SM. Also, $ M_R $ is the Majorana neutrino mass matrix 
 and $ C $ denotes the charge-conjugation operator. After spontaneous symmetry breaking,  one obtains the Dirac neutrino  mass term as $\overline{\nu}_{\alpha L}^{} M_D^{} N_R^{} + \mathrm{h.c.}$, where $M_D^{} = v Y_\nu^{}$ is the  Dirac neutrino mass matrix with  vacuum expectation value ($ vev $), $v = \langle H \rangle \approx 174~\mathrm{GeV}$ \cite{Patrignani:2016xqp}.
Employing seesaw mechanism, one gets light neutrino mass matrix in type-I seesaw formalism as, $  M_{\nu}^{} \approx - M_D^{} M^{-1}_R M^{T}_D $ and diagonalization of such $ M_{\nu}^{}  $ leads to three active neutrino masses $ m_i$ (for $ i = 1, 2, 3 $).   

Furthermore, flavor symmetry based approaches get numerous attention  
to  explain the observed neutrino mixing patterns as discussed in Refs.
\cite{Altarelli:2010gt, Altarelli:2012ss,Smirnov:2011jv,Ishimori:2010au, King:2013eh} and the references therein. 
Among number of such approaches, $\mu$-$\tau$ reflection symmetry attracts a lot of attention in recent times which was originally discussed in Ref.~\cite{Harrison:2002et} (see Ref.~\cite{Xing:2015fdg} for a latest review). This symmetry  predicts : the maximal atmospheric mixing angle  $\theta_{23}^{}$, i.e., $\theta_{23}^{} = 45^\circ$ along with the maximal value of Dirac CP phase $\delta$, i.e. $\delta = \pm 90^\circ$; and trivial values for the two Majorana phases with non-zero  $\theta_{13}^{}$.
Indeed, in recent times there are many attempts toward  $\mu$-$\tau$ reflection symmetry as outlined in Refs.~\cite{Ferreira:2012ri,Mohapatra:2012tb, Ma:2015gka,Ma:2015fpa,He:2015xha, Joshipura:2015dsa, Joshipura:2015zla, Joshipura:2016hvn, Nishi:2016wki,Zhao:2017yvw,Rodejohann:2017lre,Liu:2017frs,Xing:2017mkx,Xing:2017cwb, Nath:2018hjx,Zhao:2018vxy, Chakraborty:2018dew,Nath:2018xih}. 
%

In this work, we embed $\mu$-$\tau$ reflection symmetry in minimal type-I seesaw formalism such that one can address both neutrino masses and mixing patterns (see Ref.~\cite{Guo:2006qa} for recent review). Later, we study its consequences considering next-generation super beam Deep Underground Neutrino Experiment (DUNE). This  statistically high potential experiment will improve the precision of the atmospheric mixing angle, $ \theta_{23} $  and play a key role  to probe the leptonic CP-violating phase, $ \delta $ \cite{Acciarri:2015uup}. Because of this, DUNE can test various flavor symmetry models and helps us to understand some inherent  physics associate with it.

 At the given framework along with maximal $ \delta$ and $\theta_{23}^{}$, we also find remaining oscillation parameters both analytically  as well as numerically.
Considering this as a true benchmark point, we depict the allowed area in the ($ \delta-\sin^2\theta_{23}^{}$) plane for DUNE at various confidence levels which serves  our intention to inspect precision of these two less known parameters. This also show the potential of DUNE to know  how well it can measure $ \delta$ and $ \theta_{23}^{}$.
Moreover, latest results of global-fit of neutrino oscillation data from NuFit-3.2 collaboration ~\cite{nufit18, Esteban:2016qun} favors higher octant of $\theta_{23}^{}$ along with non-maximal $ \delta$ \footnote{Note that $\theta_{23} <  45^\circ$ is called a lower octant (LO) where $\theta_{23} > 45^\circ$ is called a higher octant (HO).}. Also, results of ongoing neutrino oscillation experiments (e.g., T2K~\cite{Abe:2017uxa} and NO$\nu$A~\cite{NOVA2018}) are in well agreement with the predictions of the concerned symmetry but, still show large uncertainties in their measurement of $ \delta $ and $\theta_{23}^{}$. Therefore, it is tenacious to accept the exact nature of $ \mu-\tau $ reflection symmetry. In that respect, it is worthwhile to study various  broken scenarios of such a symmetry. To proceed with phenomenological study, we first  perform our analysis considering global best-fit values as our benchmark point~\cite{nufit18, Esteban:2016qun}. Afterwards, we consider  breaking of $\mu$-$\tau$ reflection symmetry by introducing explicit breaking parameter in the high-energy neutrino mass matrices $ M_D $,  $ M_R $, respectively. For each scenario, we find the set of neutrino oscillation parameters and perform the capability test of DUNE in the ($ \delta-\sin^2\theta_{23}^{}$) plane. Considering different cases, we analyze the potential of DUNE to rule out the possibility of  maximal CP-violation (CPV) as well as CP-conservation hypothesis at a given confidence level.  
%
%
%
Some recent studies considering different flavor models in the context of long baseline experiments have been performed  in~\cite{Toorop:2011jn,Hanlon:2013ska,Hanlon:2014bga,Srivastava:2017sno,Agarwalla:2017wct,Chatterjee:2017ilf,Pasquini:2018udd,Srivastava:2018ser,Petcov:2018snn,Chakraborty:2018dew}.

We organize rest of the  paper as follows. In Sec.~\ref{sec:MuTauSymm}, we present a general setup of the $\mu-\tau$ reflection symmetry and perform our analysis in the given scenario for DUNE. We also present our numerical details in this section. We proceed to discuss our results considering NuFit-3.2 data in Sec.~\ref{sec:BrSymm}. Furthermore, in subsequent subsections of  Sec.~\ref{sec:BrSymm}, we discuss the breaking of $\mu-\tau$ reflection symmetry by introducing explicit breaking parameter in $M_D^{}$ and $M_R^{}$, respectively and their implications in the context of DUNE. Finally, we summarize our noteworthy results in Sec.~\ref{sec:conclusion}.
\section{Phenomenology at $ \mu-\tau $ reflection symmetry}\label{sec:MuTauSymm}
The $ \mu-\tau $ reflection symmetry at the low-energy neutrino mass matrix, $ M_{\nu} $ was first proposed in  Ref.~\cite{Harrison:2002et} which leads us to following four predictions :
\begin{eqnarray} \label{eq:Mnu_pred}
M_{ee}^{} = M_{ee}^* \; , \quad M_{\mu\tau}^{} = M_{\mu\tau}^* \; , \quad M_{e\mu}^{} = M_{e\tau}^* \; , \quad M_{\mu\mu}^{} = M_{\tau\tau}^* \; ,
\end{eqnarray}
where $ M_{\alpha \beta}, ({\rm with } ~\alpha,\beta = e, \mu, \tau )$ are the elements of  $ M_{\nu} $.
We consider minimal type-I seesaw mechanism to realize $ \mu-\tau $ reflection symmetry at $ M_{\nu} $. To achieve such symmetry, we extend the SM fields content by adding two right-handed neutrino fields which are singlet under the SM gauge group. Without loss of generality, we consider the following texture of $ M_D $  to realize $ \mu-\tau $ reflection symmetry,
\begin{eqnarray}\label{eq:md}
M_D = \left( \begin{matrix} a & a^{\ast} \cr
 b & c \cr  c^{\ast} & b^{\ast}
\end{matrix} \right)   = \left( \begin{matrix} a e^{ i \phi_{a}}   & a e^{- i \phi_{a}} \cr
 b e^{ i \phi_{b}} & c e^{ i \phi_{c}} \cr  c e^{- i \phi_{c}} & b  e^{-i \phi_{b}}
\end{matrix} \right)  \;. \end{eqnarray} 
Also, we adopt diagonal  $ M_R $ of the form  $ M_R = diag (M_1, M_1) $  with degenerate heavy Majorana neutrino masses \footnote{It is possible to find the considered mass textures using a suitable flavor group along with preferred $\mathbb{Z}_n$ cyclic group. As our intention is to study the impact of these textures rather their theoretical origin, hence we do not perform this study here.}.
Further, considering type-I seesaw mechanism, we obtain the effective neutrino mass matrix for the light neutrinos as
%
\begin{eqnarray}\label{eq:MnuSymm}
-M_{\nu} & = & M_D M^{-1}_R M^{T}_D ,  \nonumber \\
%
 & = & \dfrac{1}{M_1} \left(
 \begin{array}{ccc}
 2 a^2 \cos2 \text{$\phi_a$} &  a b^{} e^{i (\text{$\phi_a$}+\text{$\phi_b $})} +a c e^{- i (\text{$\phi_a $}-\text{$\phi_c $})}  & a b e^{-i (\text{$\phi_a $}+\text{$\phi_b $})} + a c e^{i (\text{$\phi_a $}-\text{$\phi_c $})} \\
- & b^2 e^{2 i \text{$\phi_b $}} +c^2 e^{2 i \text{$\phi_c$}} & 2 b c \cos (\text{$\phi_b$}-\text{$\phi_c$}) \\
- & - & b^2 e^{-2 i \text{$\phi_b$}} + c^2 e^{-2 i \text{$\phi_c$}} \\
\end{array}
\right) \;.
\end{eqnarray}
We notice that the elements of $ M_{\nu} $ as given by Eq.~(\ref{eq:MnuSymm}) satisfy all the conditions of Eq.~(\ref{eq:Mnu_pred}) and hence leads to $ \mu-\tau $ reflection 
symmetry \footnote{Note that non-degenerate Majorana neutrino mass matrix does not satisfy all the conditions mentioned in Eq.~(\ref{eq:Mnu_pred}) and thus does not lead to the concerned symmetry  which we discuss in section~\ref{sec:BrSymm}.}.
In the standard PDG \cite{Patrignani:2016xqp} parametrization, the unitary mixing matrix which diagonalizes neutrino mass matrix, $  M_{\nu}  $,  can be written as,
\begin{align}\label{eq:pmns}
V & = P_l U P_{\nu}, \nonumber \\
 & = P_l \left( \begin{matrix}
c^{}_{12} c^{}_{13} & s^{}_{12} c^{}_{13} & s^{}_{13} e^{-{\rm i} \delta} \cr 
 -s^{}_{12} c^{}_{23} - c^{}_{12} s^{}_{13} s^{}_{23} e^{{\rm i} \delta} & c^{}_{12} c^{}_{23} -
s^{}_{12} s^{}_{13} s^{}_{23} e^{{\rm i} \delta} & c^{}_{13}
s^{}_{23} \cr 
s^{}_{12} s^{}_{23} - c^{}_{12} s^{}_{13} c^{}_{23}
e^{{\rm i} \delta} & -c^{}_{12} s^{}_{23} - s^{}_{12} s^{}_{13}
c^{}_{23} e^{{\rm i} \delta} &  c^{}_{13} c^{}_{23} \cr
\end{matrix} \right)  P_{\nu}, \;
\end{align}
where $c^{}_{ij} = \cos\theta^{}_{ij}, s^{}_{ij} = \sin\theta^{}_{ij}$ (for $i<j=1,2,3$). Here,  $ P_l $ contains three unphysical phases of the form $ P_l = diag(e^{i \phi_{e}},e^{i \phi_{\mu}},e^{i \phi_{\tau}}) $ which can be absorbed by the rephasing of charged lepton fields,  whereas $ P_{\nu} = diag(e^{i \rho},e^{i \sigma},1) $ contains two Majorana phases.

With the above form of $ M_{\nu}  $ as given by Eq.~(\ref{eq:MnuSymm}), one can find that there exist 6 predictions for the leptonic mixing angles and phases  which are \footnote{For a detailed discussion on the adopted phase conventions see appendix of Ref.~\cite{Nath:2018hjx}.},
\begin{equation}\label{eq:prediction}
\phi_{e} = 90,~~~ \phi_{\mu} = - \phi_{\tau}=\phi,~~~ \theta_{23} = 45^\circ,~~~ \delta=\pm 90^\circ,~~~ \rho,~\sigma = 0^\circ ~~{\rm or}~~ 90^\circ.
\end{equation}
Note that under $ \mu-\tau $ reflection symmetry  the value of $ \theta_{13} $, $  \theta_{12}$ remain unspecified. We find their analytical form in terms of model parameters as \footnote{Note that mass pattern of the form $ m_3 > m_2> m_1 $ is known as normal mass ordering (NMO) whereas  $ m_3 < m_1 \approx m_2 $ pattern is known as inverted mass ordering (IMO).},
\begin{align}\label{eq:AtMuTuSymm}
 \theta_{13} & = \mp \tan^{-1}\left[ \dfrac{b^2 \sin 2\varphi_b+c^2 \sin 2\varphi_c}{a (b \sin \varphi_{ab} +c \sin\varphi_{ac}) } \right]  \;, \nonumber \\[10pt]
\theta^{}_{12} & = 
\begin{cases}
\dfrac{1}{2}\tan^{-1} \left[ \dfrac{2\sqrt{2}a \cos 2\theta_{13} ( b \sin \varphi_{ab} +c \sin  \varphi_{ac}) }
 {\splitfrac{c_{13}[(   b^2 \cos2\varphi_b+c^2 \cos2\varphi_c   - 2 b c \cos\varphi_{bc}  )  \cos 2\theta_{13} }
{ - (  b^2 \cos2\varphi_b+c^2 \cos2\varphi_c +  2 b c \cos\varphi_{bc}  )   s^{2}_{13} + 2 a^2 \cos 2\phi_a  c^{2}_{13}  ] } } \right]   ~; ~~{\rm for }  ~~ {\rm NMO }  \;  \\[30pt]
\dfrac{1}{2}\tan^{-1} \left[ \dfrac{2\sqrt{2} a( b \sin \varphi_{ab} +c \sin\varphi_{ac} )  s^{2}_{13} }{c_{13}\left[ ( b^2 \cos2\varphi_b+c^2 \cos2\varphi_c ) (1 + s^{2}_{13}) + 2  c^{2}_{13} b c \cos\varphi_{bc} \right] } \right] ~; ~~{\rm for }  ~~ {\rm IMO }
\end{cases}
\end{align}
where $ \varphi_{b,c}  =  ( \phi - \phi_{b,c}),~\varphi_{ab,c}  = (\phi -\phi_a-\phi_{b,c}), ~  \varphi_{bc}  = - ( \phi_b - \phi_c)$.

Similarly, one can calculate masses of light neutrinos by diagonalizing $  M_{\nu}$ of Eq.(\ref{eq:MnuSymm})  as
\begin{align}
V^{\dagger}M_{\nu}V^{*} & =  diag(m_1,m_2,m_3) .
\end{align}
where $ m_i$'s $( i = 1,2,3)$ are the active neutrino masses.
Further, the masses can be expressed for NMO as 
\begin{eqnarray}
m_1 & = &0 \;, \nonumber \\
\widehat{m}_2 & = &\dfrac{2\sqrt{2} a (b \sin \varphi_{ab} +c \sin\varphi_{ac}) }{c_{13}\sin 2 \theta_{12} M_1}\;, \nonumber \\
m_3 & = & \dfrac{1}{M_1} \left[  4 b c \cos\varphi_{bc} + 2 a^2 \cos2\phi_a +  \dfrac{2\sqrt{2}  a (b \sin \varphi_{ab} +c \sin\varphi_{ac})  }{c_{13}\sin 2 \theta_{12}}  \right] \; ,
\end{eqnarray}
 whereas, expressions for  IMO can be written as
\begin{eqnarray}
m_1 & = & \dfrac{1}{M_1}  \left[ 2 b c \cos\varphi_{bc}  - a^2 \cos2\phi_a -  \dfrac{2\sqrt{2}  a (b \sin \varphi_{ab} +c \sin\varphi_{ac})  }{c_{13}\sin 2 \theta_{12}} \right]   \;,\nonumber \\
\widehat{m}_2 & = & \dfrac{1}{M_1}  \left[ - 2 b c \cos\varphi_{bc}  - a^2 \cos2\phi_a +  \dfrac{2\sqrt{2}  a (b \sin \varphi_{ab} +c \sin\varphi_{ac})  }{c_{13}\sin 2 \theta_{12}} \right]  \;, \nonumber \\
m_3 & = & 0 \; .
\end{eqnarray}
Here, $ \widehat{m}_2 = m_2^{} e^{2i\sigma} $  and $ \sigma $ can take value either $ 0^\circ $ or $ 90^\circ $. Also note that as the minimal seesaw formalism always predicts massless lightest neutrino, one has the freedom of eliminating one of the Majorana phases. Thus, in this study we do not consider phase, $ \rho $.

To proceed further and to investigate low-energy phenomenology, we first give here simulation and experimental details that are considered in this work. 
The principle strategy of our numerical analysis is to scan all the high-energy variables of  $ Y_{\nu}$ and  $M_R^{}$ as free variables and later constrain the allowed space of high-energy variables to find neutrino oscillation parameters  which are compatible with the latest NuFit-3.2 data ~\cite{nufit18, Esteban:2016qun} at low energies. We vary different parameters as, 
\begin{eqnarray}\label{eq:Variables}
|a|, |b|, |c| \in [0, 1]~v, ~~~\phi_{a,b,c} \in [0, 360^\circ), ~ ~~  \quad M_1 \in [10^{12}, 10^{15}]~\mathrm{GeV} \;.
\end{eqnarray}
We use the nested sampling package $\texttt{Multinest}$ \cite{Feroz:2007kg,Feroz:2008xx,Feroz:2013hea} to guide the parameter scan with the built $\chi^2$ function considering latest NuFit-3.2 data~\cite{nufit18, Esteban:2016qun}. 
The analytical expression of the Gaussian-$\chi^2_{\rm min}$ function that we use in our numerical simulation is defined as,
\begin{equation}\label{eq:ChiSqMin}
\chi^{2}_{ min} ={\rm min} \sum_i \dfrac{\left[  \xi_i^{\rm True} - \xi_i^{\rm Test} \right] ^{2}  }{\sigma \left[ \xi_i^{\rm True} \right] ^{2}} \;,
\end{equation}
where  $\xi = \{ \theta_{12}, \theta_{13} , \theta_{23} , \Delta m_{21}^{2}, |\Delta m_{31}^{2}| \}$,  represents the set of neutrino oscillation parameters. Here, $\xi_i^{\mathrm{Ture}}$ represent the current best-fit values of the latest NuFit-3.2 data ~\cite{nufit18, Esteban:2016qun} and $\xi_i^{\mathrm{Test}}$ correspond to the predicted values for a given set of parameters in theory. We also  symmetrize standard deviation, $\sigma \left[ \xi_i^{\rm True} \right]$ considering 1$\sigma$ errors as given by Ref.~\cite{nufit18, Esteban:2016qun}.

We consider here DUNE, which is a proposed next generation superbeam experiment at Fermilab, USA 
\cite{Acciarri:2015uup, Alion:2016uaj} designing to detect neutrinos.
This experiment will utilize existing NuMI (Neutrinos at the Main Injector) 
beamline design at Fermilab as a neutrino source. The far detector of DUNE will be placed at Sanford Underground Research Facility (SURF) in Lead, South Dakota, at a distance of 1300 km (800 mile)
from neutrino source. DUNE collaboration has planned to use LArTPC 
(liquid argon time-projection chamber) detector.
%
For the numerical simulation of the DUNE data, we use the
\texttt{GLoBES} package \cite{globes1, globes2} along 
with the required auxiliary files presented in Ref.~\cite{Alion:2016uaj}.
We perform our simulation  considering 40 kton fiducial mass far detector. 
We also consider the flux corresponding to 1.07 MW beam power which gives
 $1.47\times 10^{21} $ protons on target (POT) per year due to 80 GeV proton beam energy. 
In addition, we adopt signal and background normalization uncertainties for appearance as well as disappearance channel as presented in DUNE CDR~\cite{Alion:2016uaj}. Further,  we  distribute the total exposure of DUNE (i.e., 300 kton-MW-years) in two scenarios ; (i) in the first scenario, we perform our analysis only with neutrino mode considering 7 years of neutrino run, i.e.,  DUNE[$ 7\nu + 0\overline{\nu} $], and (ii)  in the second scenario, we consider 3.5 years each of the neutrino and antineutrino mode i.e., DUNE[$ 3.5\nu + 3.5\overline{\nu} $].
 We also add $5\%$ prior on sin$^{2}2\theta_{13}$ in our analysis.

The main steps to carry out our numerical analysis are to calculate set of neutrino oscillation parameters corresponding to the minimum $ \chi^{2} ( = \chi^{2}_{\rm min} )$, as defined by Eq.(\ref{eq:ChiSqMin}), using $\texttt{Multinest}$ in this model. Later, considering this set of parameters as true benchmark value, we generate DUNE results using \texttt{GLoBES} and present the allowed parameter space in the test ($ \delta - \sin^2\theta_{23} $) plane. We utilize the \texttt{GLoBES} inbuilt $  \chi^{2} $-function for the data analysis. In this study, we marginalize all the oscillation parameters over their 3$ \sigma $ range as given by Table~\ref{tab:NuFit}. In addition, we marginalize $ \delta $ in the range $ \delta \in [0^\circ, 360^\circ )$ for each scenario unless otherwise stated.

\begin{figure}[!h]
\begin{center}
 \begin{tabular}{lr}
\hspace{-1.5cm}
\includegraphics[height=13cm,width=14cm]{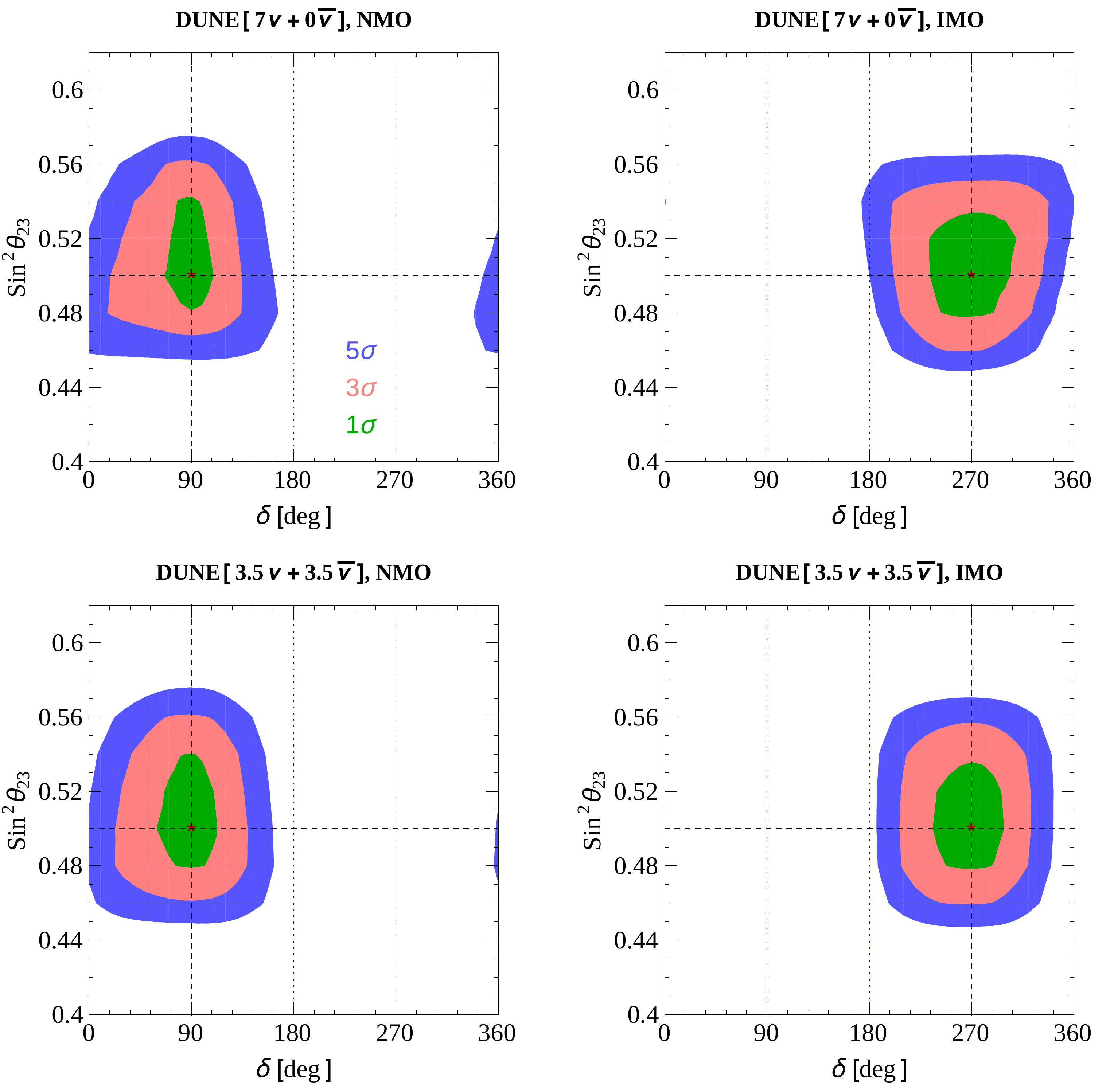}
 \end{tabular}
 \end{center}
\vspace{-4ex}        
\caption{\footnotesize Allowed parameter space of DUNE in the ($ \delta - \sin^2\theta_{23} $) plane in the $\mu-\tau$ reflection symmetry scenario. Here, the green, pink, and blue color represent 1$ \sigma, 3\sigma $, and 5$ \sigma $ allowed contours and `red-$\ast$' signifies the true value of  ($ \delta, \sin^2\theta_{23} $). Also the left (right) column represents normal (inverted) mass ordering and the top (bottom) row shows our results for DUNE[$7\nu + 0\overline{\nu}$] ( DUNE[$3.5\nu + 3.5\overline{\nu}]$ ).}
\label{fig:MuTauSymm}
\end{figure}

 \begin{table}[htb]
        \centering \scriptsize
       \begin{tabular}{|c|c|c| }
       \hline  Parameters    &  NMO &    IMO  \\ 
                    &  ($ \chi^2_{min} = 0.10 $) &  ($ \chi^2_{min} = 0.82$) \\
       \hline $\Delta \text{m}^{2}_{21} [10^{-5} {\rm eV}^{2}] $  &  7.401  & 7.50  \\ 
       \hline $|\Delta \text{m}^{2}_{31}|[10^{-3} {\rm eV}^{2}] $  & 2.498 &  2.465 \\ 
       \hline $ \sin^{2}\theta_{12} $ & 0.304  & 0.303   \\ 
       \hline  $\sin^{2}\theta_{23}$ & 0.50 &   0.50    \\
       \hline  $\sin^{2}\theta_{13} $ & 0.02217  & 0.02218  \\
       \hline $\delta $ [deg] & 90 & 270 \\
       \hline
     \end{tabular}
     \caption{\footnotesize Set of neutrino oscillation parameters at $ \chi^{2}_{\rm min} = 0.10$  ($ \chi^{2}_{\rm min} = 0.82$ ) for NMO (IMO) in  the $\mu-\tau$ reflection symmetry scenario.} \label{tab:Symm}       
      \end{table} 
In Fig.\ref{fig:MuTauSymm}, we present our results in the framework of $ \mu - \tau $ reflection symmetry. We calculate the numerical values for the set of neutrino oscillation parameters in the given scenario corresponding to $  \chi^2_{min}$ as given in Table~\ref{tab:Symm}. Considering this true set of parameters, we find  the allowed area in the ($ \delta - \sin^2\theta_{23} $) plane in case of DUNE which we have depicted in Fig.\ref{fig:MuTauSymm}. 
The  green-, pink-, and blue-colored contours represent 1$ \sigma, 3\sigma $, and 5$ \sigma $ allowed parameter space, respectively  and the red-star point represents the true value of  ($ \delta, \sin^2\theta_{23} $). Further, the top and bottom row show our results for DUNE[$7\nu + 0\overline{\nu}$] and DUNE[$3.5\nu + 3.5\overline{\nu}$], respectively \footnote{Note that authors of Ref.~\cite{Nath:2015kjg} have performed a detailed analysis on the sensitivity of these less known parameters considering various combinations of ($\nu + \overline{\nu}$) for DUNE.}.  Also,  the vertical black-dashed lines represent maximal CPV corresponding to $ \delta = 90^\circ $ and $ 270^\circ$,  respectively. Similarly, the blue-dotted line signifies  the CP-conserving value $ \delta = 180^\circ $, and horizontal black-dashed line represents $ \sin^2\theta_{23}  = 0.5$.  Note that we consider similar color details throughout this work.
 
Considering maximal value of  ($ \delta, \sin^2\theta_{23} $) as a true benchmark point, we notice from the first row of Fig.\ref{fig:MuTauSymm} that 
7 years of neutrino run of DUNE can rule out CP-conservation hypothesis at 1$ \sigma $ C.L. for both the mass ordering (i.e., NMO, IMO) as shown by green contour. This observation remains true even at  3$ \sigma $ C.L. for both the mass ordering as presented in the pink contour.  To justify this point, we notice from upper panel that the pink contour does not intersect with  vertical blue-dotted line which provides clear evidence of the ruling out of CP-conservation hypothesis at the same confidence level.  Besides this,  we notice from the 5$ \sigma $ contour (see blue contour) that DUNE cannot exclude CP-conservation hypothesis for both the mass orderings. In addition, we also notice that the precision of CP-phase, $ \delta $ is marginally better in the case of IMO compared to NMO,  where $ \sin^2\theta_{23} $ shows almost similar precision for both cases at 5$ \sigma $ C.L. From the second row of Fig.\ref{fig:MuTauSymm}, we notice that DUNE can rule out the CP-conservation hypothesis even at 5$ \sigma $ C.L. for IMO (see right panel) where in the case NMO, it can almost exclude the same except for some regions around ($ \delta = 0^\circ/360^\circ,  \sin^2\theta_{23} = 0.5$).
 Finally, we notice from the top row that DUNE can rule out one half-plane of $ \delta $ at  3$ \sigma $ C.L., but at 5$ \sigma $ C.L. it can exclude almost the same for both the mass orderings. 
In the case of NMO (for true $ \delta  = 90^\circ$ ), we observe that DUNE can rule out $ \delta $ in the range $\delta \in [180^\circ, 360^\circ ]$ whereas for IMO  (for true $ \delta  = 270^\circ$ ), it can rule out $ \delta $ in the range, $\delta \in [0^\circ, 180^\circ ]$  at 3$ \sigma $ C.L. 
Similarly, from the bottom row we notice that the same conclusion remains true even at 5$ \sigma $ C.L. except for small regions for NMO. 
 
Having discussed our results in the  $ \mu - \tau $ reflection symmetry scenario considering DUNE, in the following section we proceed to perform our analysis by utilizing current oscillations data. Later, we  also examine different symmetry breaking scenarios where we will discuss the impact of breaking parameter on the poorly measured parameters, $ \delta$ and $ \sin^2\theta_{23} $.
\section{Phenomenology Beyond $\mu-\tau$ Reflection Symmetry}\label{sec:BrSymm}
In this section, we discuss our results beyond $\mu-\tau$ reflection symmetry considering DUNE. 
As the current best-fit value of neutrino oscillation data  prefers non-maximal value of $ \delta$, $ \sin^2\theta_{23} $, we start the discussion considering  this as our true  benchmark point. Furthermore, in subsequent subsections, we perform our study considering different  breaking scenarios  of  $\mu-\tau$ reflection symmetry and its impact in the context of DUNE.
\subsection{Analysis of global best-fit data}
In Table~\ref{tab:NuFit}, we give the latest results of global-fit of neutrino oscillation data as obtained by NuFit-3.2 \cite{nufit18} collaboration. We notice from the table that the best-fit points of latest analysis favor higher octant for the 2-3 mixing angle, $\theta_{23}$ and non-maximal value for the Dirac CP phase, $ \delta $ for both the mass orderings.

\begin{figure}[!h]
\begin{center}
 \begin{tabular}{lr}
\hspace{-1.5cm}
\includegraphics[height=13cm,width=14cm]{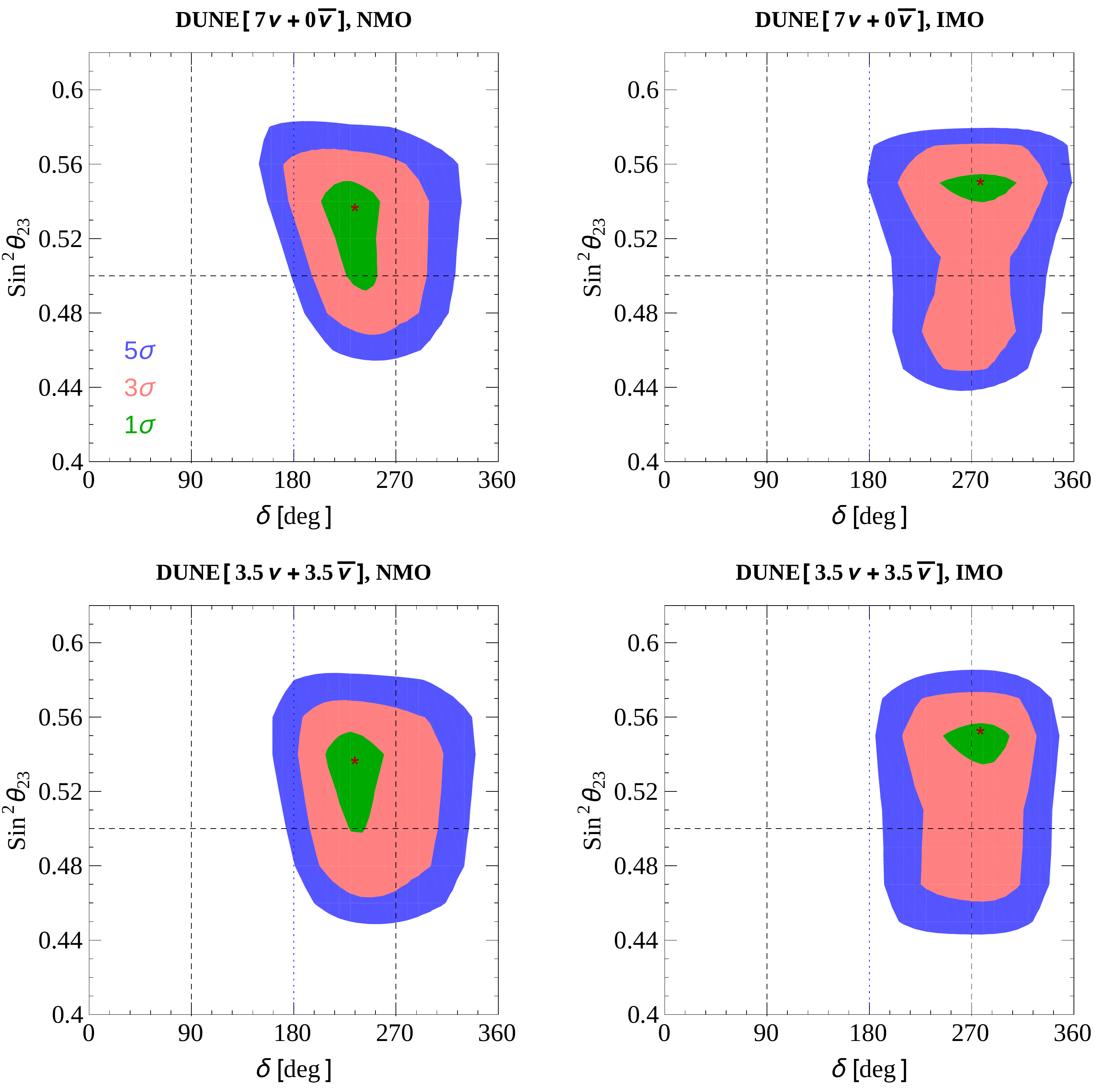}
 \end{tabular}
 \end{center}
\vspace{-4ex}        
\caption{\footnotesize Allowed parameter space of DUNE in the ($ \delta - \sin^2\theta_{23} $)-plane considering latest NuFit-3.2 data \cite{nufit18}. Remaining details are the same as Fig.~\ref{fig:MuTauSymm}.}
\label{fig:GlobalFit}
\end{figure}

 \begin{table}[htb]
        \centering \scriptsize
       \begin{tabular}{|c|c|c|c| }
       \hline  Oscillation    &  NMO &    IMO & Any Ordering \\ 
                Parameters    & Best-fit & Best-fit & 3$ \sigma $ \\
       \hline $\Delta \text{m}^{2}_{21} [10^{-5} {\rm eV}^{2}] $  &  7.40  & 7.40 & 6.80 $ \rightarrow $ 8.02 \\ 
       \hline $|\Delta \text{m}^{2}_{31}|[10^{-3} {\rm eV}^{2}] $  & 2.494 &  2.465 & 2.399 $ \rightarrow $ 2.593 (NMO)\\ 
       & &  &  2.395 $ \rightarrow $ 2.536 (IMO)\\
       \hline $ \sin^{2}\theta_{12} $ & 0.307  & 0.307 & 0.272 $ \rightarrow $ 0.346  \\ 
       \hline  $\sin^{2}\theta_{23}$ & 0.538 &   0.554 & 0.418 $ \rightarrow $ 0.613   \\
       \hline  $\sin^{2}\theta_{13} $ & 0.02206  & 0.02227 & 0.019  $ \rightarrow $ 0.0243 \\
       \hline $\delta $ [deg] & 234 & 278 & 144 $ \rightarrow $ 374\\
       \hline
     \end{tabular}
     \caption{\footnotesize The best-fit values and 3$ \sigma $ range of neutrino oscillation parameters \cite{nufit18}} \label{tab:NuFit}       
      \end{table} 
In Fig.~\ref{fig:GlobalFit}, we present our results  in the ($ \delta - \sin^2\theta_{23} $)-plane for DUNE considering best-fit values of NuFit-3.2 data as our true benchmark point. 
Here red-star represents true value of ($ \delta, \sin^2\theta_{23} $) i.e., ($ 234^\circ, 0.538 $) and ($ 278^\circ, 0.554 $) corresponding to NMO and IMO, respectively. 
From first plot of top row, we notice that DUNE can exclude the possibility of having maximal CP-violation as well as CP-conservation hypothesis at 1$ \sigma $ C.L. as shown by the green contour for NMO.
On the other hand, it cannot exclude either of these hypotheses at 3$ \sigma $ C.L. as can be seen from the pink contour which intersects with $ \delta = 180^\circ $ vertical blue-dotted line and  $ \delta = 270^\circ $ vertical black-dashed line. 
%
Investigating bottom row for normal mass ordering, we notice from first plot that DUNE can exclude maximal CP-violation at 1$ \sigma $ C.L. similar as only neutrino mode of DUNE. Apart from this it can exclude CP-conservation hypothesis at 3$ \sigma $ C.L. but not at higher confidence levels.

In the case of IMO, as shown in the right column, we notice that DUNE cannot exclude the phenomenon of maximal CP-violation even at 1$ \sigma $ C.L. as depicted by green contour.  But, it can exclude  CP-conservation hypothesis approximately at 5$ \sigma $ C.L. as the blue contour marginally  touches $ \delta = 180^\circ $. 
On the other hand,   it can reject the CP-conservation hypothesis  even at 5$ \sigma $ C.L. as described by the blue contour of last plot for inverted mass ordering with 3.5 years each of neutrino and antineutrino run of DUNE. Furthermore,  normal mass ordering of DUNE[$ 3.5\nu + 3.5\overline{\nu} $] can marginally exclude  lower octant (LO) of $ \theta_{23} $ (i.e., when $ \sin^2\theta_{23} \leq 0.5 $) at 1$ \sigma $ C.L. as depicted by green contour.
Moreover, in the case of IMO, we notice that it can rule out LO of $ \theta_{23} $  clearly at 1$ \sigma $ C.L. but not at higher confidence levels.
\subsection{Breaking of $\mu-\tau$ reflection symmetry through $ M_D $}
We discuss here three different scenarios to break $\mu-\tau$ reflection symmetry by introducing explicit breaking parameter in the Dirac neutrino mass matrix,  $ M_D $. Further, for each case we perform  precision study to determine $ \delta, \sin^2\theta_{23} $  considering DUNE.  We study them as follows:
\begin{itemize}
\item Broken Scenario-1 (\textbf{BS1}): \label{sec:BS1}
After assigning a breaking parameter in the (12) position of  $ M_D $, the new Dirac neutrino mass matrix,  $  \widehat{M}_D $ can be written as
\begin{eqnarray}\label{eq:BrMd12}
 \widehat{M}_D  = \left( \begin{matrix} a e^{ i \phi_{a}}   & a (1 + \epsilon) e^{- i \phi_{a}} \cr
 b e^{ i \phi_{b}} & c e^{ i \phi_{c}} \cr  c e^{- i \phi_{c}} & b  e^{-i \phi_{b}}
\end{matrix} \right)  \;. \end{eqnarray} 

The above texture of $  \widehat{M}_D $ leads to low-energy neutrino mass matrix $  \widehat{M}_{\nu} $ of the form,
\begin{eqnarray}\label{eq:MnuMd12} 
 \widehat{M}_{\nu} & \simeq &  M_{\nu} - \epsilon \dfrac{a e^{- i \phi_a } }{M_1} ~ 
\left(
\begin{array}{ccc}
 2 a e^{- i \phi_a } &  c e^{i \phi_c}   &  b e^{-i  \phi_b}   \\
  c e^{ i \phi_c }   & 0 & 0 \\
  b e^{- i  \phi_b}   & 0 & 0 \\
\end{array}
\right) 
+ \mathcal{O}(\epsilon^{2})\;.
\end{eqnarray}
Now to find masses and mixing angles in presence of breaking term $ \epsilon $, we diagonalize 
$  \widehat{M}_{\nu} $ with $  \widehat{V}$ \footnote{We vary the breaking term $ \epsilon $ in the range, [-1,1] along with other high-energy parameters, as mentioned in Eq.~(\ref{eq:Variables}).}. Note that $\widehat{V}$ has similar form as $V$ in the absence of $ \epsilon $ as described by Eq.~(\ref{eq:pmns}). In Table~\ref{tab:MdBr12}, we give the expressions of modified masses and mixing angles for both the mass orderings. Note that for simplicity, we only consider the leading order corrections in terms of $ \epsilon $, $ \theta_{13} $ and $ \xi_{1}= m_2/m_3 (\xi_{2} = \Delta m^{2}_{21}/m^{2}_2)$ for NMO (IMO).

\begin{table}[!h]
\centering \scriptsize
\begin{tabular}{ c || c | c }
\hline \hline Parameters(\textbf{S1})  &  NMO &    IMO \\ 
\hline 
\hline
$\widehat{m}_1  \simeq $  &  0  &  $\begin{aligned}[t] m_1 - & \epsilon \dfrac{a}{M_1} [2 a c^{2}_{12} \cos2\phi_{a} 
\\ &  + \sqrt{2} s_{12}c_{12} (b \sin\varphi^{\mu}_{ab} - c \sin\varphi^{\mu}_{ac} )] \end{aligned}$ \\ 
$\widehat{m}_2  \simeq $  & $\begin{aligned}[t] m_2 - & \epsilon \dfrac{a}{M_1} [2 a s^{2}_{12} \cos2\phi_{a} 
\\ &  + \sqrt{2} s_{12}c_{12} (c \sin\varphi^{\mu}_{ac} - b \sin\varphi^{\mu}_{ab} )] \end{aligned}$  &  $\begin{aligned}[t] m_2 - & \epsilon \dfrac{a}{M_1} [2 a s^{2}_{12} \cos2\phi_{a} 
\\ &  + \sqrt{2} s_{12}c_{12} (c \sin\varphi^{\mu}_{ac} - b \sin\varphi^{\mu}_{ab} )] \end{aligned}$ \\[25pt] 
$\widehat{m}_3  \simeq $  & $ m_3 - \epsilon \dfrac{\sqrt{2} a }{M_1}  [b\cos\varphi^{\mu}_{ab} + c \cos\varphi^{\mu}_{ac}]\theta_{13}  $   &  0 \\[7pt]
\hline 
$ \widehat{\theta}_{13} \simeq  $ & $ \theta_{13} -  \epsilon \dfrac{a}{\sqrt{2}m_3 M_1} [b\cos\varphi^{\mu}_{ab} + c \cos\varphi^{\mu}_{ac}] $ & $ \theta_{13} +  \epsilon \dfrac{a}{\sqrt{2}m_2 M_1} [b\cos\varphi^{\mu}_{ab} + c \cos\varphi^{\mu}_{ac}] $\\[10pt]
$ \widehat{\theta}_{12} \simeq  $ & $ \begin{aligned}[t]
 \theta_{12} - &   \epsilon \dfrac{a}{2 m_3 M_1 \xi_{1}} [2a\cos2\phi_{a} \sin2\theta_{12} \\ & + \sqrt{2}\cos2\theta_{12} (c \sin\varphi^{\mu}_{ac} - b\sin\varphi^{\mu}_{ab})]  \end{aligned} $ &
$ \begin{aligned}[t]
 \theta_{12} - &   \epsilon \dfrac{a}{m_2 M_1 \xi_{2}} [2a\cos2\phi_{a} \sin2\theta_{12} \\ & + \sqrt{2}\cos2\theta_{12} (c \sin\varphi^{\mu}_{ac} - b\sin\varphi^{\mu}_{ab})]  \end{aligned} $
  \\[25pt]
$ \widehat{\theta}_{23} \simeq  $ & $ 45^\circ + \epsilon \dfrac{a}{\sqrt{2}m_3 M_1} [b\cos\varphi^{\mu}_{ab} - c \cos\varphi^{\mu}_{ac}]\theta_{13}  $ & $ 45^\circ + \epsilon \dfrac{a}{\sqrt{2}m_2 M_1} [b\cos\varphi^{\mu}_{ab} - c \cos\varphi^{\mu}_{ac}]\theta_{13}  $\\[10pt]
\hline
\hline
\end{tabular}
\caption{\footnotesize Modified masses and mixing angles in \textbf{BS1} scenario. Here, the second (third) column represents expressions for NMO (INO). Also, we used notation $ \varphi^{\mu}_{ac}=  (\phi_a- \phi_c +\phi_\mu ) $, $ \varphi^{\mu}_{ab}=  (\phi_a + \phi_b-\phi_\mu ) $.}
\label{tab:MdBr12}       
\end{table} 
Afterwards, we proceed to find the set of neutrino oscillation parameters numerically in this scenario. We also emphasize here that  the numerical analysis throughout this work are based on exact formula not on any leading order approximations. The numerical best-fit values at  $ \chi^2_{min} $ for both the mass orderings are tabulated in Table \ref{tab:NumMdBr12}.  Considering this set of values as the true benchmark point, we present allowed area in the test ($ \delta - \sin^2\theta_{23} $)-plane for DUNE  in Fig.~\ref{fig:Md12} \footnote{Note that one can also perform various correlations study considering neutrino oscillation parameters in different broken scenarios. Recently, authors in Ref.~\cite{Nath:2018hjx}  have performed different correlation study. }.

 \begin{table}[htb]
        \centering \scriptsize
       \begin{tabular}{|c|c|c| }
       \hline  Parameters    &  NMO &    IMO  \\ 
                    &  ($ \chi^2_{min} = 0.71 $) &  ($ \chi^2_{min} = 4.5$) \\
       \hline $\Delta \text{m}^{2}_{21} [10^{-5} {\rm eV}^{2}] $  &  7.34  & 7.56  \\ 
       \hline $|\Delta \text{m}^{2}_{31}|[10^{-3} {\rm eV}^{2}] $  & 2.49 &  2.47 \\ 
       \hline $ \sin^{2}\theta_{12} $ & 0.312  & 0.310   \\ 
       \hline  $\sin^{2}\theta_{23}$ & 0.514 &   0.499    \\
       \hline  $\sin^{2}\theta_{13} $ & 0.02235  & 0.02228  \\
       \hline $\delta $ [deg] & 350 & 65 \\
       \hline
     \end{tabular}
     \caption{\footnotesize Set of neutrino oscillation parameters corresponding to $ \chi^{2}_{\rm min} = 0.71 ( = 4.5 ) $ for NMO (IMO) in the \textbf{BS1} scenario.} 
     \label{tab:NumMdBr12}       
      \end{table} 

\begin{figure}[!h]
\begin{center}
 \begin{tabular}{lr}
\hspace{-1.5cm}
\includegraphics[height=13cm,width=14cm]{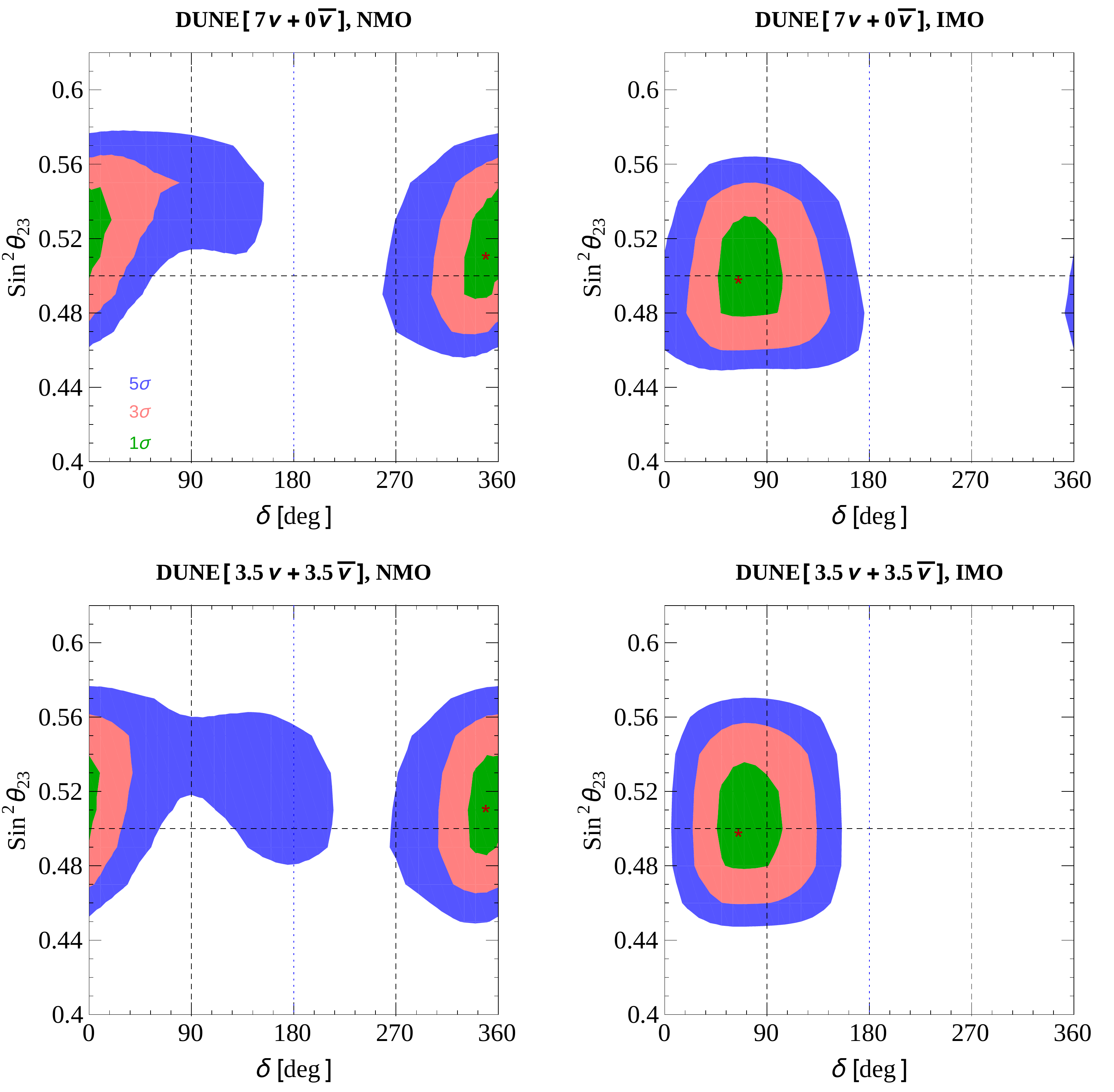}
 \end{tabular}
 \end{center}
\vspace{-4ex}        
\caption{\footnotesize Allowed parameter space of DUNE in the ($ \delta - \sin^2\theta_{23} $) plane in the \textbf{BS1} scenario. Here green, pink and blue color represent 1$ \sigma, 3\sigma $, and 5$ \sigma $ allowed contours and `red-$\ast$'  signifies  true value of  ($ \delta, \sin^2\theta_{23} $).}
\label{fig:Md12}
\end{figure}
%
From first plot of Fig.~\ref{fig:Md12}, we notice that in the \textbf{BS1} scenario, DUNE can exclude 
the theory of maximal CPV at 3$ \sigma $ C.L. (see pink contour) for NMO even only with neutrino run. On the other hand at 5$ \sigma $ C.L., this case is unable to exclude both the concerned hypotheses in neutrino mode. 
With the combined equal neutrino and antineutrino mode analysis of  DUNE, we observe that it can exclude the possibility of maximal CPV hypothesis at 3$ \sigma $ C.L. whereas CP-precision becomes poorer at 5$ \sigma $ C.L. as shown in the first plot of the second row.
 We also notice from both the plots of first column that as the best-fit value of $ \delta $ is marginally away from CP conserving value (i.e. $ \delta = 360^\circ $), this scenario cannot exclude CP-conservation hypothesis even at 1$ \sigma $ C.L. In the case of IMO, considering the best-fit values as given by third column of Table \ref{tab:NumMdBr12} as the benchmark point, we notice that DUNE can  exclude the phenomenon of CP-conservation at 3$ \sigma $ C.L. but not at 5$ \sigma $ C.L. which is depicted in first plot of second column by the pink contour.
From second plot of right column, we observe that DUNE can reject CP-conservation hypothesis even at 5$ \sigma $ C.L. Further, both the cases of IMO cannot reject the value corresponding to maximal CPV even at 1$ \sigma $ C.L.
We also notice that  precision of $ \delta $ improves significantly when one chooses IMO over NMO and it gets even better with the combined mode of DUNE run  as shown in the last plot. Finally, here we point out that DUNE can exclude $ \delta $ in the range, $\delta \in [180^\circ, 360^\circ ]$ at 3$ \sigma $ C.L. for IMO (see first plot of right column), whereas the same conclusion remains permissible even at 5$ \sigma $ C.L. with combined ($ \nu + \overline{\nu} $) analysis of DUNE (see second plot of right column).
\item Broken Scenario-2 (\textbf{BS2}): 
In this scenario, we introduce the breaking term $ \epsilon $ in the (22) position of  $ M_D$ and this modifies $ M_D$ (which we renamed $  \widehat{M}_D $)  as
\begin{eqnarray}\label{eq:BrMd22}
 \widehat{M}_D  = \left( \begin{matrix} a e^{ i \phi_{a}}   & a  e^{- i \phi_{a}} \cr
 b e^{ i \phi_{b}} & c (1 + \epsilon) e^{ i \phi_{c}} \cr  c e^{- i \phi_{c}} & b  e^{-i \phi_{b}}
\end{matrix} \right)  \;. \end{eqnarray} 

Using the form  of $  \widehat{M}_D $ as given by Eq.~(\ref{eq:BrMd22}), we find modified $  \widehat{M}_{\nu} $ as,
\begin{eqnarray}\label{eq:MnuMd22}
 \widehat{M}_{\nu} & \simeq &  M_{\nu} - \epsilon \dfrac{c e^{ i \phi_{c}}}{M_1} ~ 
\left(
\begin{array}{ccc}
 0 & a  e^{- i  \phi_{a}}  & 0 \\
 a  e^{- i \phi_{a}}  & 2 c e^{ i \phi_{c}}  & b  e^{-i \phi_{b}}  \\
 0 & b  e^{-i  \phi_{b}}  & 0 \\
\end{array}
\right) 
+ \mathcal{O}(\epsilon^{2})\;.
\end{eqnarray}
To find modified masses and mixing angles in the given scenario, we follow the similar steps as discussed in subsection \ref{sec:BS1}. In the following Table \ref{tab:MdBr22}, we give their expressions for both the mass orderings. The subleading order term in $ \epsilon $ shows the corrections in active neutrino masses  and mixing angles for this broken pattern.

\begin{table}[!h]
\centering \scriptsize
\begin{tabular}{ c || c | c  }
\hline \hline Parameters(\textbf{S2})   &  NMO &    IMO \\ 
\hline \hline
$\widehat{m}_1  \simeq $  &  0  & $\begin{aligned}[t] m_1 + & \epsilon \dfrac{c}{M_1} [\sqrt{2} a s^2_{12}\sin\varphi^{\mu}_{ac} 
\\ &  + b s_{12}c_{12} \sin\varphi^{\mu}_{ac} + s^2_{12}\cos2(\phi_{c} - \phi_{\mu})] \end{aligned}$   \\ 
$\widehat{m}_2  \simeq $  & $\begin{aligned}[t] m_2 + & \epsilon \dfrac{c}{M_1} [- \sqrt{2} a s_{12}c_{12} \sin\varphi^{\mu}_{ac} 
\\ &  + c^2_{12} (c \cos2(\phi_{c} - \phi_{\mu}) - b \cos\varphi_{bc} )] \end{aligned}$  
&  $\begin{aligned}[t] m_2 + & \epsilon \dfrac{c}{M_1} [- \sqrt{2} a s_{12}c_{12} \sin\varphi^{\mu}_{ac} 
\\ &  + c^2_{12} (c \cos2(\phi_{c} - \phi_{\mu}) - b \cos\varphi_{bc} )] \end{aligned}$   \\[25pt] 
$\widehat{m}_3  \simeq $  & $ m_3 - \epsilon \dfrac{c }{M_1}  [b\cos\varphi_{bc} + c \cos2(\phi_{c} - \phi_{\mu})] $   &  0 \\[7pt]
\hline 
$ \widehat{\theta}_{13} \simeq  $ & $ \theta_{13} -  \epsilon \dfrac{a c}{\sqrt{2}m_3 M_1} \cos\varphi^{\mu}_{ac} $ &  $ \theta_{13} + \epsilon \dfrac{a c}{\sqrt{2}m_2 M_1} \cos\varphi^{\mu}_{ac} $\\[10pt]
$ \widehat{\theta}_{12} \simeq  $ & $ \begin{aligned}[t]
 \theta_{12} - &   \epsilon \dfrac{c}{2 m_3 M_1 \xi_{1}} [c( \cos2(\phi_{c} - \phi_{\mu} ) - b \cos\phi_{bc}) \\ & \times \sin2\theta_{12}  + \sqrt{2} a\cos2\theta_{12} \sin\varphi^{\mu}_{ac}]  \end{aligned} $ &
$ \begin{aligned}[t]
 \theta_{12} - &   \epsilon \dfrac{c}{ m_2 M_1 \xi_{2}} [c( \cos2(\phi_{c} - \phi_{\mu} ) - b \cos\phi_{bc}) \\ & \times \sin2\theta_{12} + \sqrt{2} a\cos2\theta_{12} \sin\varphi^{\mu}_{ac}]  \end{aligned} $
  \\[25pt]
$ \widehat{\theta}_{23} \simeq  $ & $ 45^\circ - \epsilon \dfrac{c^2}{m_3 M_1} \cos2(\phi_{c}-\phi_{\mu})   $ &  $ 45^\circ - \epsilon \dfrac{c^2}{m_2 M_1} \cos2(\phi_{c}-\phi_{\mu})   $\\[10pt]
\hline \hline
\end{tabular}
\caption{\footnotesize Modified masses and mixing angles in the \textbf{BS2} scenario. 
Notation adopted here are same as Eq.~(\ref{eq:AtMuTuSymm}) and Table~\ref{tab:MdBr12}.
}
\label{tab:MdBr22}       
\end{table} 
After finding analytical expressions, we now evaluate the numerical set of neutrino oscillation parameters in the broken scenario \textbf{BS2}. We calculate the best-fit values corresponding to $ \chi^2_{min} $ numerically and present them in Table \ref{tab:NumBrMd22}. 

 \begin{table}[htb]
        \centering \scriptsize
       \begin{tabular}{|c|c|c| }
       \hline  Parameters    &  NMO &    IMO  \\ 
                    &  ($ \chi^2_{min} = 1.01 $) &  ($ \chi^2_{min} = 4.58$) \\
       \hline $\Delta \text{m}^{2}_{21} [10^{-5} {\rm eV}^{2}] $  &  7.428  & 7.56  \\ 
       \hline $|\Delta \text{m}^{2}_{31}|[10^{-3} {\rm eV}^{2}] $  & 2.499 &  2.450 \\ 
       \hline $ \sin^{2}\theta_{12} $ & 0.305  & 0.301   \\ 
       \hline  $\sin^{2}\theta_{23}$ & 0.49 &   0.51   \\
       \hline  $\sin^{2}\theta_{13} $ & 0.0218  & 0.0229  \\
       \hline $\delta $ [deg] & 89 & 125 \\
       \hline
     \end{tabular}
     \caption{\footnotesize  Set of neutrino oscillation parameters corresponding to $ \chi^{2}_{\rm min} = 1.01 ( = 4.58 ) $ for NMO (IMO) in the \textbf{BS2} scenario.} 
     \label{tab:NumBrMd22}       
      \end{table} 

\begin{figure}[!h]
\begin{center}
 \begin{tabular}{lr}
\hspace{-1.5cm}
\includegraphics[height=13cm,width=14cm]{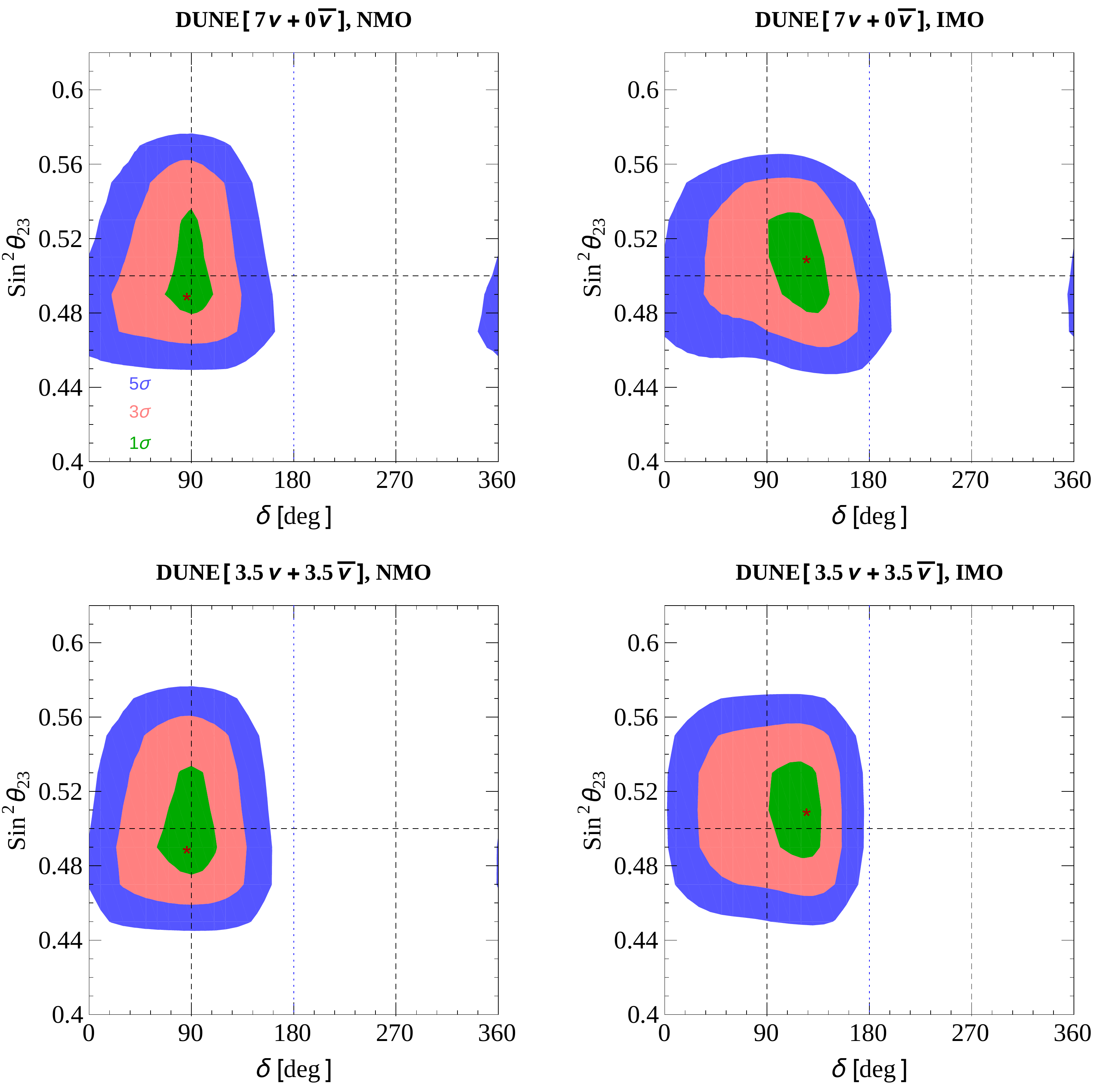}
 \end{tabular}
 \end{center}
\vspace{-4ex}        
\caption{\footnotesize Allowed parameter space of DUNE in the ($ \delta - \sin^2\theta_{23} $) plane in the  \textbf{BS2} scenario. Here green, pink, and blue colors represent 1$ \sigma, 3\sigma $ and 5$ \sigma $ allowed contours and red-$ \ast $ signifies  true value of  ($ \delta, \sin^2\theta_{23} $).}
\label{fig:Md22}
\end{figure}
Moreover, considering the given set of values as our true benchmark point, we show allowed parameter space of DUNE in the ($ \delta - \sin^2\theta_{23} $)-plane for NMO as well as IMO in Fig.~\ref{fig:Md22}. In case of NMO, as described in left panel, we notice that as the concerned scenario predicts, $ \delta = 89^\circ$ and $\sin^2\theta_{23} = 0.49 $ neutrino mode of DUNE cannot rule out maximal CPV hypothesis even at 1$ \sigma $ C.L. This observation prolongs things further, even with the inclusion of the antineutrino run with the neutrino as depicted by the second plot of the first panel, whereas the neutrino mode of it can exclude  the CP-conservation hypothesis at 3$ \sigma $. In addition, the combined ($ \nu + \overline{\nu} $) run can reject the same scenario approximately at 5$ \sigma $ C.L. as shown by the blue contour. We also notice by comparing both the plots of the left column that DUNE can reject $ \delta $ in the range, $\delta \in [180^\circ, 360^\circ ]$ at 3$ \sigma $,  5$ \sigma $ C.L. considering only neutrino and  combined ($ \nu + \overline{\nu} $) mode run of DUNE, respectively.
 From right panel (which is for IMO), we find that both cases can  rule out maximal CPV as well as CP-conservation hypothesis  only at 1$ \sigma $ C.L. 
Besides this, we notice that  only neutrino mode data of DUNE can exclude the CP-conservation hypothesis at 3$ \sigma $ C.L. but not at 5$ \sigma $ C.L., whereas the combined effect of ($ \nu + \overline{\nu} $) can reject the same hypothesis at 5$ \sigma $ C.L. as shown in the bottom right panel by the blue contour. 
At the end, we notice from the right panel that CP precision  improves significantly with the combined effect of the neutrino and antineutrino run for  DUNE and it can successfully  exclude $ \delta $ in the range $\delta \in [180^\circ, 360^\circ ]$ at 5$ \sigma $ C.L. In addition, comparing both the columns we find here that NMO shows better CP precision over IMO.
\item Broken Scenario-3 (\textbf{BS3}): 
Here, we  assign the breaking parameter in the (32) position of  $ M_D $,  and we write the new Dirac neutrino mass matrix,  $  \widehat{M}_D $  as
\begin{eqnarray}\label{eq:BrMd32}
 \widehat{M}_D  = \left( \begin{matrix} a e^{ i \phi_{a}}   & a  e^{- i \phi_{a}} \cr
 b e^{ i \phi_{b}} & c e^{ i \phi_{c}} \cr  c e^{- i \phi_{c}} & b (1 + \epsilon) e^{-i \phi_{b}}
\end{matrix} \right)  \;. \end{eqnarray} 

$  \widehat{M}_D $ given by Eq.~(\ref{eq:BrMd32}), leads us to the following $  \widehat{M}_{\nu}$ through type-I seesaw formalism,
\begin{eqnarray}\label{eq:MnuMd32}
 \widehat{M}_{\nu} & \simeq &  M_{\nu} - \epsilon \dfrac{b e^{- i  \phi_{b}}}{M_1} ~ 
\left(
\begin{array}{ccc}
 0 & 0 & a  e^{-i  \phi_{a}  } \\
 0 & 0 &  c e^{i  \phi_{c} } \\
  b e^{-i \phi_{a}  } &  c e^{i \phi_{c}} & 2 b e^{-2 i  \phi_{b}} \\
\end{array}
\right)
+ \mathcal{O}(\epsilon^{2})\;.
\end{eqnarray}
Now we diagonalize  $  \widehat{M}_{\nu} $ as given by Eq.~(\ref{eq:MnuMd32}) to find corrections in masses and mixing angles. Here also we perform  similar study as discussed in subsection \ref{sec:BS1}. In  Table \ref{tab:MdBr32}, we give analytical expressions for masses and mixing angles considering both the mass orderings where  $\mathcal{O}(\epsilon) $ term  shows the correction in active neutrino masses  and mixing angles for the concerned scenario.


\begin{table}[!h]
\centering \scriptsize
\begin{tabular}{ c || c | c  }
\hline\hline  Parameters(\textbf{S3})   &  NMO &    IMO \\ 
\hline \hline
$\widehat{m}_1  \simeq $  &  0  &  $\begin{aligned}[t] m_2 + & \epsilon \dfrac{b}{M_1} [- \sqrt{2} a s_{12}c_{12} \sin\varphi^{\mu}_{ab} 
\\ &  + c^2_{12} (b \cos2(\phi_{b} - \phi_{\mu}) - c \cos\varphi_{bc} )] \end{aligned}$   \\ 
$\widehat{m}_2  \simeq $  & $\begin{aligned}[t] m_2 + & \epsilon \dfrac{b}{M_1} [ \sqrt{2} a s_{12}c_{12} \sin\varphi^{\mu}_{ab} 
\\ &  + c^2_{12} (b \cos2(\phi_{b} - \phi_{\mu}) - c \cos\varphi_{bc} )] \end{aligned}$  
&  $\begin{aligned}[t] m_2 + & \epsilon \dfrac{b}{M_1} [ \sqrt{2} a s_{12}c_{12} \sin\varphi^{\mu}_{ab} 
\\ &  + c^2_{12} (b \cos2(\phi_{b} - \phi_{\mu}) - c \cos\varphi_{bc} )] \end{aligned}$   \\[25pt] 
$\widehat{m}_3  \simeq $  & $ m_3 - \epsilon \dfrac{b }{M_1}  [b  \cos2(\phi_{b} - \phi_{\mu}) + c\cos\varphi_{bc} ] $   &  0 \\[7pt]
\hline 
$ \widehat{\theta}_{13} \simeq  $ & $ \theta_{13} -  \epsilon \dfrac{a b}{\sqrt{2}m_3 M_1} \cos\varphi^{\mu}_{ab} $ & $ \theta_{13} +  \epsilon \dfrac{ab}{\sqrt{2}m_2 M_1} \cos\varphi^{\mu}_{ab}  $\\[10pt]
$ \widehat{\theta}_{12} \simeq  $ & $ \begin{aligned}[t]
 \theta_{12} + &   \epsilon \dfrac{b}{2 m_3 M_1 \xi_{1}} [( c \cos\phi_{bc} - b \cos2(\phi_{b} - \phi_{\mu} ) ) \sin2\theta_{12} \\ & + \sqrt{2} a\cos2\theta_{12} \sin\varphi^{\mu}_{ab}]  \end{aligned} $ &
$ \begin{aligned}[t]
 \theta_{12} + &   \epsilon \dfrac{b}{m_2 M_1 \xi_{2}} [( c \cos\phi_{bc} - b \cos2(\phi_{b} - \phi_{\mu} ) ) \sin2\theta_{12} \\ & + \sqrt{2} a\cos2\theta_{12} \sin\varphi^{\mu}_{ab}]  \end{aligned} $
  \\[25pt]
$ \widehat{\theta}_{23} \simeq  $ & $ 45^\circ + \epsilon \dfrac{b^2}{m_3 M_1} \cos2(\phi_{b}-\phi_{\mu})   $ &  $ 45^\circ + \epsilon \dfrac{b^2}{m_2 M_1} \cos2(\phi_{b}-\phi_{\mu})   $\\[10pt]
\hline\hline
\end{tabular}
\caption{\footnotesize Modified masses and mixing angles in the \textbf{BS3} scenario. 
Notation adopted here are same as Eq.~(\ref{eq:AtMuTuSymm}) and Table~\ref{tab:MdBr12}.
}
\label{tab:MdBr32}       
\end{table} 

 \begin{table}[htb]
        \centering \scriptsize
       \begin{tabular}{|c|c|c| }
       \hline  Parameters    &  NMO &    IMO  \\ 
                    &  ($ \chi^2_{min} = 0.62 $) &  ($ \chi^2_{min} = 5.39$) \\
       \hline $\Delta \text{m}^{2}_{21} [10^{-5} {\rm eV}^{2}] $  &  7.49  & 7.28  \\ 
       \hline $|\Delta \text{m}^{2}_{31}|[10^{-3} {\rm eV}^{2}] $  & 2.493 &  2.428 \\ 
       \hline $ \sin^{2}\theta_{12} $ & 0.311  & 0.316  \\ 
       \hline  $\sin^{2}\theta_{23}$ & 0.56 &   0.51    \\
       \hline  $\sin^{2}\theta_{13} $ & 0.0219  & 0.0229  \\
       \hline $\delta $ [deg] & 252 & 140 \\
       \hline
     \end{tabular}
     \caption{\footnotesize  Set of neutrino oscillation parameters corresponding to $ \chi^{2}_{\rm min} = 0.62 ~( = 5.39 ) $ for NMO (IMO) in \textbf{BS3} scenario.} 
     \label{tab:NumBrMd32}       
      \end{table} 
Having discussed analytical results, we proceed to find the set of neutrino oscillation parameters in the broken scenario \textbf{BS3}. We calculate the best-fit values corresponding to $ \chi^2_{min} $ numerically and present them in Table \ref{tab:NumBrMd32}. Using this set of true benchmark points, we examine allowed parameter space of DUNE in the ($ \delta - \sin^2\theta_{23} $) plane for both the mass orderings as shown in Fig.~\ref{fig:Md32} (see figure caption for the adopted color convention and other minutes details).

\begin{figure}[!h]
\begin{center}
 \begin{tabular}{lr}
\hspace{-1.5cm}
\includegraphics[height=13cm,width=14cm]{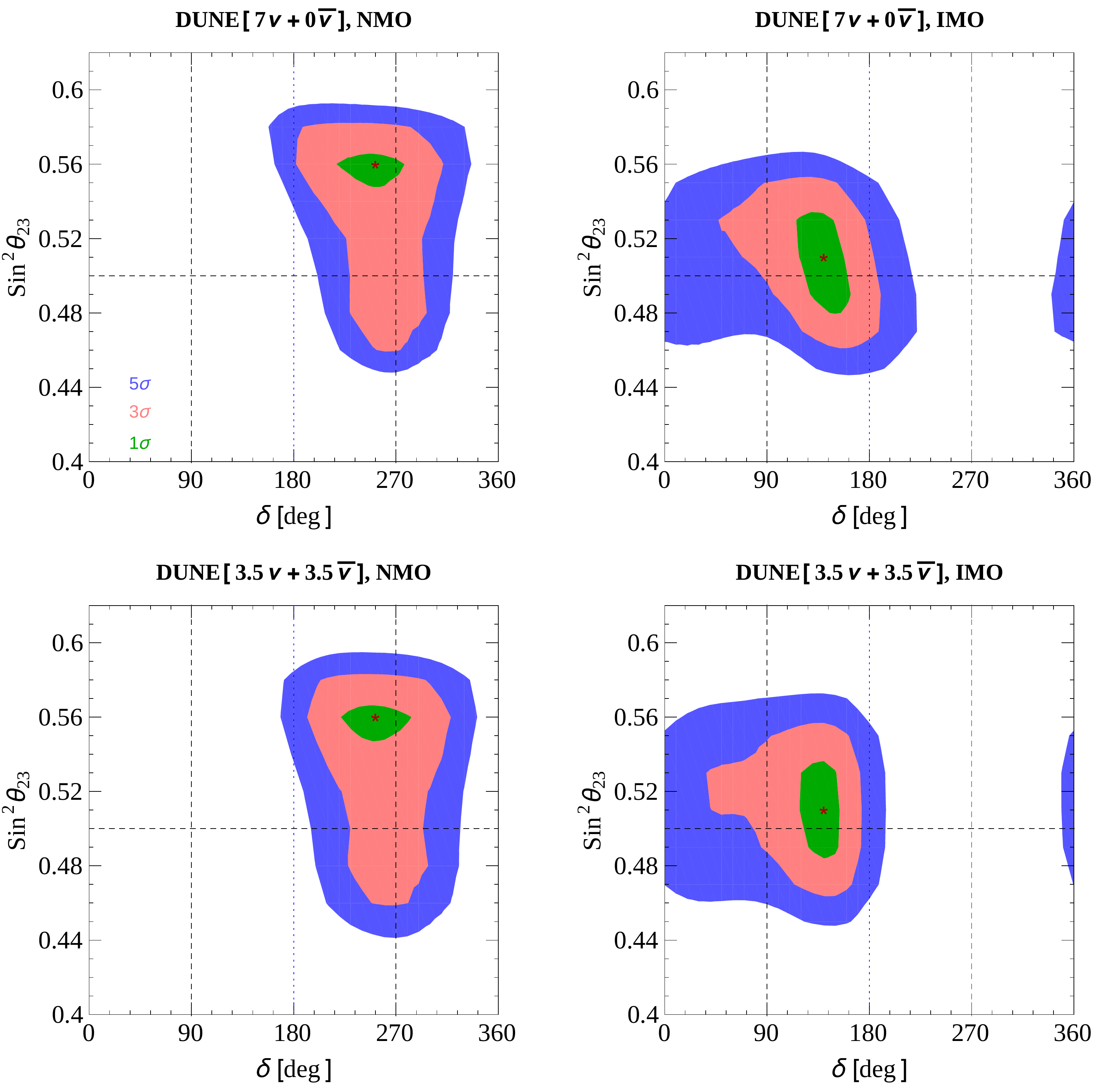}
 \end{tabular}
 \end{center}
\vspace{-4ex}        
\caption{\footnotesize Allowed parameter space of DUNE in the ($ \delta - \sin^2\theta_{23} $) plane in  the \textbf{BS3} scenario. Here, green, pink, and blue colors represent 1$ \sigma, 3\sigma $ and 5$ \sigma $ allowed contours and red-$ \ast $ signifies  true value of  ($ \delta, \sin^2\theta_{23} $).}
\label{fig:Md32}
\end{figure}
%
We observe from first plot of left panel that DUNE with only neutrino mode data is not able to exclude the phenomenon of maximal CPV   even at 1$ \sigma $ C.L. (see green contour for NMO), whereas it  can exclude the CP-conservation hypothesis at 3 $ \sigma $ C.L. (see pink contour for NMO) but not at 5$ \sigma $ C.L. as the blue contour intersect with the blue-dotted vertical line. We find that similar conclusion remains permissible for  the combined effect of ($ \nu + \overline{\nu} $) run of DUNE as shown in first plot of second row. 
  In the case of IMO with 7-years neutrino run, we find that DUNE can reject both the concerned hypotheses at 1$ \sigma $ C.L. At higher confidence levels, however,  it fails to rule out any of these hypotheses as depicted in the first plot of the second panel.
Investegeting the right hand side plot of second row, we notice that at 1$ \sigma $ C.L. it shows similar behaviour as neutrino mode whereas at 3$ \sigma $ C.L. it is able to rule out CP-conservation hypothesis but not maximal CPV as shown by the pink contour.
Finally, we observe a noteworthy outcome in this scenario compared to the former two breaking patterns this scenario can exclude the lower octant of $ \theta_{23} $  at 1$ \sigma $ C.L. for NMO even with 7-years of neutrino mode data of DUNE.
\end{itemize}
\subsection{Breaking of $\mu-\tau$ reflection symmetry through $ M_R $}
We discuss here the  breaking of $\mu-\tau$ reflection symmetry by introducing explicit breaking parameter in the Majorana neutrino mass matrix, $ M_R $.
We discuss  the scenario as below.
\begin{itemize}
\item Broken Scenario-4 (\textbf{BS4}): 
After assigning the breaking parameter in the (22) position of  $ M_R $, the modified Majorana neutrino mass matrix,  $  \widehat{M}_R $ becomes,
\begin{eqnarray}\label{eq:BrMr22}
 \widehat{M}_R  = \left( \begin{matrix} M_1   & 0 \cr
0 & M_1 (1 + \epsilon) 
\end{matrix} \right)  \;. \end{eqnarray} 
Note here that in this scenario $  \widehat{M}_R $ becomes non-degenerate.
After integrating out heavy right-handed neutrino fields, the low-energy neutrino mass matrix in the type-I seesaw formalism can be written as
\begin{eqnarray}\label{eq:MnuMr22}
 \widehat{M}_{\nu} & \simeq &  M_{\nu} -  \dfrac{\epsilon}{M_1} ~ 
\left(
\begin{array}{ccc}
 a^2 e^{-2 i  \phi_{a}} & a c^{} e^{- i (\phi_{a} - \phi_{c})} & a^{} b^{} e^{- i (\phi_{a} + \phi_{b})} \\
 - &  b^2 e^{-2 i  \phi_{b}} &  b^{} c^{} e^{- i (\phi_{b} - \phi_{c})} \\
- &  - &  c^2 e^{-2 i  \phi_{c}} \\
\end{array}
\right)
+ \mathcal{O}(\epsilon^{2})\;.
\end{eqnarray}

 \begin{table}[htb]
        \centering \scriptsize
       \begin{tabular}{|c|c|c| }
       \hline  Parameters    &  NMO &    IMO  \\ 
                    &  ($ \chi^2_{min} = 0.53 $) &  ($ \chi^2_{min} = 3.91$) \\
       \hline $\Delta \text{m}^{2}_{21} [10^{-5} {\rm eV}^{2}] $  &  7.31  & 7.38  \\ 
       \hline $|\Delta \text{m}^{2}_{31}|[10^{-3} {\rm eV}^{2}] $  & 2.497 &  2.456 \\ 
       \hline $ \sin^{2}\theta_{12} $ & 0.302  & 0.303   \\ 
       \hline  $\sin^{2}\theta_{23}$ & 0.53 &   0.50    \\
       \hline  $\sin^{2}\theta_{13} $ & 0.02179  & 0.02228  \\
       \hline $\delta $ [deg] & 280 & 33 \\
       \hline
     \end{tabular}
     \caption{\footnotesize  Set of neutrino oscillation parameters corresponding to $ \chi^{2}_{\rm min} = 0.53 ~ ( = 3.91 ) $ for NMO (IMO) in the \textbf{BS4} scenario.} 
     \label{tab:BrMR22}       
      \end{table} 
In this framework, we notice from Eq.~(\ref{eq:MnuMr22}) that as all the entries of $ \mathcal{O}(\epsilon) $ term are non-zero, it is highly non-trivial to perform analytical study and to find expressions for modified neutrino masses and mixing angles. Therefore, we proceed to employ only numerical study unlike previous subsections where both analytical as well as numerical study was performed. The set of neutrino oscillation parameters at  $ \chi^2_{min} $ for possible mass ordering is tabulated in Table \ref{tab:BrMR22}. We notice from table that best-fit values corresponding to $ \chi^2_{min} $ deviates from maximal ($ \delta, \theta_{23} $) for NMO whereas for IMO the given mass textures still favor maximal $ \theta_{23} $ but not maximal $ \delta$.

\begin{figure}[!h]
\begin{center}
 \begin{tabular}{lr}
\hspace{-1.5cm}
\includegraphics[height=13cm,width=14cm]{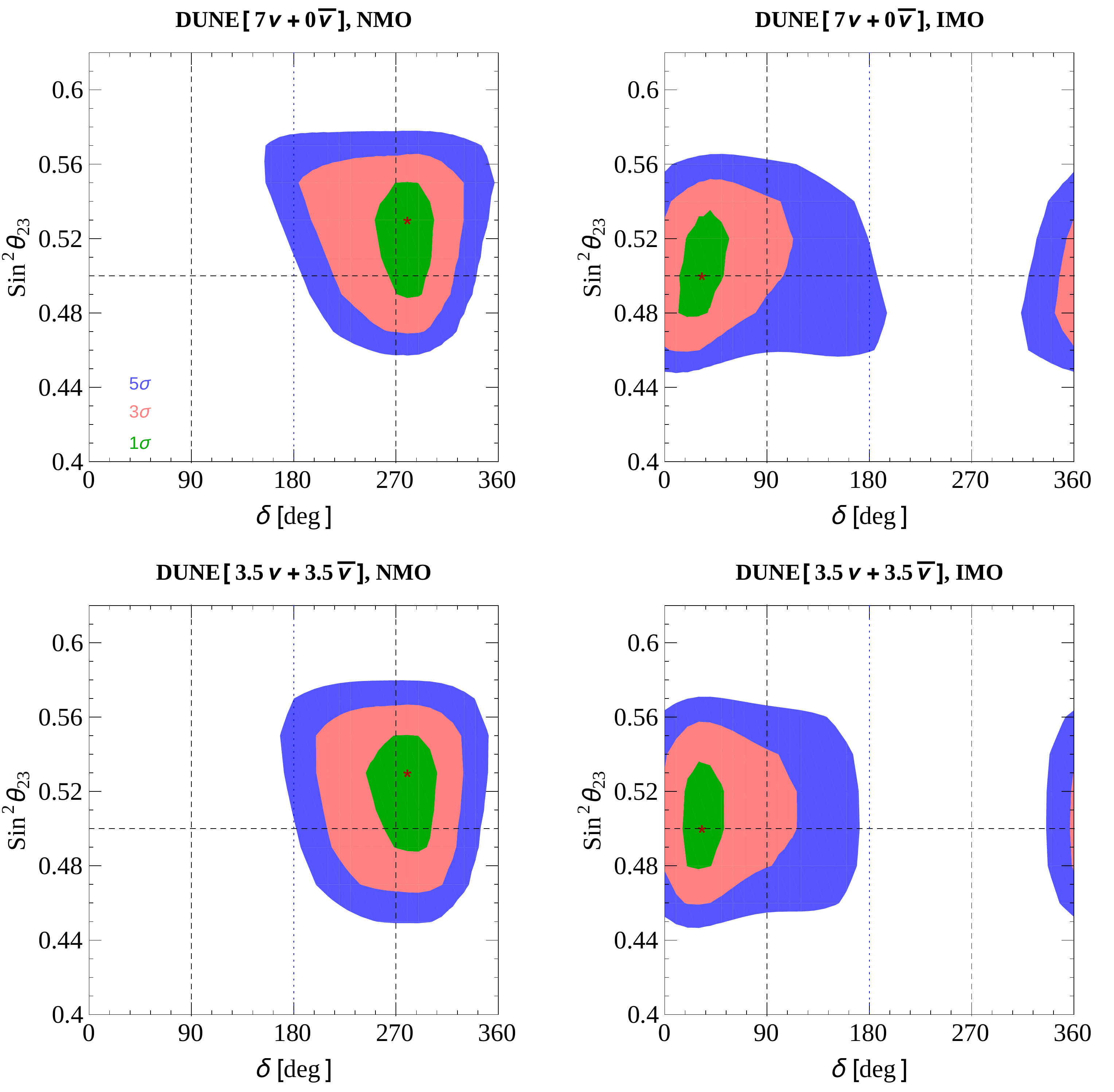}
 \end{tabular}
 \end{center}
\vspace{-4ex}        
\caption{\footnotesize Allowed parameter space of DUNE in ($ \delta - \sin^2\theta_{23} $)-plane for the  \textbf{BS4} scenario. Remaining details are same as Fig.~\ref{fig:MuTauSymm}.}
\label{fig:Mr22}
\end{figure}
After finding the set of best-fit values at $ \chi^{2}_{\rm min} $, we proceed to analyze its impact on DUNE. Performing similar kinds of analysis as illustrated in the former broken scenarios,  we also show here the allowed parameter space of DUNE considering two poorly determined parameters, viz, $ \delta$ and $\sin^2\theta_{23} $. We show our results in Fig.~\ref{fig:Mr22} considering the test ($ \delta - \sin^2\theta_{23} $) plane. Now  from both the plots of first column, we notice that as the given mass textures have chosen
the value of Dirac CP-phase, $ \delta $ slightly away from its maximal value at $ \chi^{2}_{\rm min} $,  DUNE fails to rule out the phenomenon of maximal CPV even at 1$ \sigma $ C.L. In fact, it can rule out CP-conservation hypothesis  at 3$ \sigma $ C.L. even with only neutrino run as shown in first plot of top row by pink contour. We see similar conclusion from the second plot of first column.
 We also notice here that DUNE with 3.5 years of each neutrino and antineutrino mode data can approximately exclude $ \delta $ in the range, $\delta \in [0^\circ, 180^\circ ]$  at 5$ \sigma $ C.L. for the normal mass ordering.  In the case of IMO, as depicted in the right column, we find that DUNE can exclude both the concerned phenomena, viz, maximal CPV and CP-conservation at 1$ \sigma $ C.L. but not at higher confidence levels.
Also, none of the cases  are able to rule out lower octant of $\sin^2\theta_{23} $ even at 1$ \sigma $ C.L. In addition, we find here that NMO shows better CP-precision over IMO.

We add a remark here  that as the Majorana neutrino mass matrix is always symmetric, addition of non-zero off-diagonal
entry still respect $ \mu - \tau $ flavor symmetry and predicts maximal $ \delta $, $ \sin^{2}\theta_{23} $. Hence, here  we do not include this as an additional scenario.
\end{itemize}

\begin{table}[!h]
\centering \scriptsize
\begin{tabular}{| c | c | c | c |}
\hline
\multicolumn{4}{|c|}{ mCPV (CPC) } \\
\hline
 \makecell{ Scenarios \\ NMO } &
 \makecell{ $1\sigma$ \\ ($ 7\nu + 0\overline{\nu} $) \quad  \quad  ($ 3.5\nu + 3.5\overline{\nu} $)} & \makecell{ $3 \sigma $ \\ ($ 7\nu + 0\overline{\nu} $)  \quad\quad ($ 3.5\nu + 3.5\overline{\nu} $)}  & \makecell{ $5 \sigma $ \\ ($ 7\nu + 0\overline{\nu} $) \quad \quad ($ 3.5\nu + 3.5\overline{\nu} $)}  \\[10pt]
\hline
$ \mu - \tau  $ & \xmark (\checkmark)\quad\quad\quad \quad\quad \xmark (\checkmark) & \xmark (\checkmark)\quad\quad\quad\quad\quad \xmark (\checkmark) & \xmark (\xmark)\quad\quad\quad\quad \quad \xmark (\xmark) \\[5pt]
GF & \checkmark (\checkmark)\quad\quad\quad\quad \quad \checkmark(\checkmark) & \xmark (\xmark)\quad\quad \quad\quad\quad \xmark (\checkmark) & \xmark (\xmark)\quad\quad \quad\quad\quad \xmark (\xmark) \\[5pt]
\textbf{BS1} & \checkmark (\xmark)\quad\quad\quad\quad \quad  \checkmark (\xmark) & \checkmark (\xmark) \quad\quad\quad\quad \quad  \checkmark (\xmark) & \xmark (\xmark) \quad\quad \quad\quad\quad    \xmark (\xmark)\\[5pt]
\textbf{BS2}  & \xmark (\checkmark) \quad\quad \quad\quad\quad  \xmark (\checkmark) &\xmark (\checkmark) \quad\quad \quad\quad\quad  \xmark (\checkmark) &\xmark (\xmark) \quad\quad\quad\quad\quad  \xmark (\xmark)\\[5pt]
\textbf{BS3} & \xmark (\checkmark) \quad\quad \quad \quad\quad \xmark(\checkmark) & \xmark (\checkmark) \quad\quad \quad\quad\quad  \xmark (\checkmark) & \xmark (\xmark) \quad\quad \quad\quad\quad \xmark (\xmark) \\[5pt]
\textbf{BS4} & \xmark (\checkmark) \quad\quad \quad\quad\quad  \xmark (\checkmark) & \xmark (\checkmark) \quad\quad\quad\quad \quad  \xmark (\checkmark) & \xmark (\xmark) \quad\quad \quad\quad\quad \xmark (\xmark)\\[5pt]
\hline
 \makecell{ Scenarios \\ IMO } &
 \makecell{ $1\sigma$ \\ ($ 7\nu + 0\overline{\nu} $) \quad  \quad  ($ 3.5\nu + 3.5\overline{\nu} $)} & \makecell{ $3 \sigma $ \\ ($ 7\nu + 0\overline{\nu} $)  \quad\quad ($ 3.5\nu + 3.5\overline{\nu} $)}  & \makecell{ $5 \sigma $ \\ ($ 7\nu + 0\overline{\nu} $) \quad \quad ($ 3.5\nu + 3.5\overline{\nu} $)}  \\[10pt]
\hline
$ \mu - \tau  $ & \xmark (\checkmark)\quad\quad\quad\quad \quad \xmark (\checkmark) & \xmark (\checkmark)\quad\quad \quad\quad\quad \xmark (\checkmark) & \xmark (\xmark)\quad\quad \quad\quad\quad \xmark (\checkmark)\\[5pt]
GF & \xmark (\checkmark)\quad\quad \quad\quad\quad \xmark (\checkmark) & \xmark (\checkmark)\quad\quad \quad\quad\quad \xmark (\checkmark) & \xmark (\xmark)\quad\quad\quad\quad \quad \xmark (\checkmark) \\[5pt]
\textbf{BS1} & \xmark (\checkmark) \quad\quad\quad\quad \quad  \xmark (\checkmark) &\xmark (\checkmark) \quad\quad\quad\quad \quad \xmark (\checkmark) & \xmark (\xmark) \quad\quad\quad\quad \quad  \xmark (\checkmark) \\[5pt]
\textbf{BS2} & \checkmark (\checkmark) \quad\quad \quad\quad\quad  \checkmark (\checkmark)  &\xmark (\checkmark) \quad\quad \quad\quad\quad  \xmark (\checkmark) & \xmark (\xmark) \quad\quad \quad\quad\quad  \xmark (\checkmark)\\[5pt]
\textbf{BS3} & \checkmark (\checkmark) \quad\quad \quad\quad\quad  \checkmark (\checkmark)  &\xmark (\xmark) \quad\quad\quad\quad \quad  \xmark (\checkmark) & \xmark (\xmark) \quad\quad \quad\quad\quad  \xmark (\xmark)\\[5pt]
\textbf{BS4} & \checkmark (\checkmark)  \quad\quad \quad\quad\quad \checkmark (\checkmark)&\xmark (\xmark)  \quad\quad\quad\quad \quad\checkmark (\xmark) &\xmark (\xmark)   \quad\quad\quad\quad \quad \xmark (\xmark) \\
 \hline
\end{tabular}
\caption {\footnotesize The possibility of ruling out maximal CP-violation (mCPV) or CP-conservation (CPC) hypothesis  for both the mass orderings at  different C.L. in the case of DUNE. We denote the concerned hypothesis (i.e. mCPV/CPC) by the `\checkmark' ( `\xmark' ) mark when DUNE is able (unable) to rule out the given scenario. Also, the parentheses inthe  bracket show our result for the CPC hypothesis.  Note that here ``$ \mu - \tau  $" refers to the symmetry scenario and the abbreviation ``GF" stands for the scenario corresponding to global-fit data.}
\label{tab:Summarize}
\end{table}
We now summarize our results in Table~\ref{tab:Summarize} for the different scenarios which are depicted in Figs.~\ref{fig:MuTauSymm}-\ref{fig:Mr22}. We show the possibility of ruling out maximal CP-violation (mCPV) or CP-conservation (CPC) hypothesis by check mark (\checkmark) considering DUNE. Whereas if DUNE fails to rule out a concerned hypothesis, we mark this with a cross (\xmark). Note that the parenthesis in bracket shows our results for CPC  hypothesis (see table caption for
details).

Finally, we calculate the precisions of the two poorly measured parameter $ \delta$ and $\sin^{2}\theta_{23}$. The precision ($P$) can be defined as 
\begin{align}
P(\delta) & = \dfrac{\delta_{\rm max} - \delta_{\rm min}}{360^\circ}\times 100\%  \;, \\ \nonumber
P(\sin^{2}\theta_{23}) & = \dfrac{(\sin^{2}\theta_{23})_{\rm max} - (\sin^{2}\theta_{23})_{\rm min} }{(\sin^{2}\theta_{23})_{\rm max} + (\sin^{2}\theta_{23})_{\rm min}} \times 100\% \;.
\end{align}
Here, $ {\rm max}~ ({\rm min} )$ refers to the maximum (minimum) value of the concerned parameter in a given contour. Also, we present the precision table considering 3$ \sigma $ confidence level for all the cases that we have considered here around their true values. 

\begin{table}
\centering \scriptsize
\begin{tabular}{| c | c | c | c | c | c | c | }
\hline
   & NMO (in \%)  & IMO (in \%)\\[10pt]
\hline
Scenarios &
 \makecell{ $P(\delta)$ \\ ($ 7\nu + 0\overline{\nu} $) \quad ($ 3.5\nu + 3.5\overline{\nu} $)}\quad \makecell{ $P(\sin^{2}\theta_{23})$ \\ ($ 7\nu + 0\overline{\nu} $) \quad ($ 3.5\nu + 3.5\overline{\nu} $)} &
\makecell{ $P(\delta)$ \\  ($ 7\nu + 0\overline{\nu} $) \quad ($ 3.5\nu + 3.5\overline{\nu} $)}\quad \makecell{ $P(\sin^{2}\theta_{23})$ \\ ($ 7\nu + 0\overline{\nu} $) \quad ($ 3.5\nu + 3.5\overline{\nu} $)} \\[10pt]
\hline
$ \mu - \tau  $ & 32.5 \quad\quad\quad\quad\quad 31.9 \quad\quad\quad  8.7 \quad\quad\quad\quad\quad 9.3 & 37.5\quad \quad\quad\quad\quad 32.2 \quad\quad\quad 8.7 \quad\quad\quad\quad\quad 9.1\\[5pt]
GF &  36.9 \quad\quad\quad\quad\quad 34.7 \quad\quad \quad 9.2 \quad\quad\quad\quad\quad 9.9 & ~35.0 \quad \quad\quad \quad\quad 31.6  \quad\quad\quad 11.6 \quad\quad\quad\quad\quad 10.5 \\[5pt]
\textbf{BS1} & 36.1 \quad \quad\quad\quad\quad 25.0 \quad\quad \quad 9.2 \quad\quad\quad\quad\quad 8.9  &  38.8 \quad\quad \quad\quad\quad 30.0 \quad\quad \quad 8.9 \quad\quad\quad\quad\quad 9.4\\[5pt]
\textbf{BS2} & 31.4 \quad\quad\quad\quad\quad 31.9 \quad \quad\quad 9.3 \quad\quad\quad\quad\quad 9.8 & 37.5 \quad\quad\quad\quad\quad 35.0\quad\quad\quad 8.9 \quad\quad\quad\quad\quad 8.5\\[5pt]
\textbf{BS3} & ~~34.2 \quad\quad\quad\quad\quad 32.7 \quad \quad\quad 11.5 \quad\quad\quad\quad\quad 11.3 & 40.8 \quad\quad \quad\quad\quad 36.9 \quad\quad \quad 8.9\quad \quad\quad\quad\quad 8.8\\[5pt]
\textbf{BS4} & 41.1 \quad \quad\quad\quad\quad 38.6 \quad\quad\quad 9.0  \quad\quad\quad\quad\quad 9.3 & 37.5 \quad\quad\quad\quad\quad 33.8 \quad\quad\quad 8.9 \quad\quad\quad\quad\quad 9.4 \\[5pt]
 \hline
\end{tabular}
\caption {\footnotesize Precision table of $ \delta, \sin^{2}\theta_{23}$ for all the considered scenarios of ($ \delta, \sin^{2}\theta_{23}$) in case of DUNE[$ 7\nu + 0\overline{\nu} $] and DUNE[$ 3.5\nu + 3.5\overline{\nu} $] at 3$ \sigma $ C.L.}
\label{tab:Precision}
\end{table}

\section{Conclusion}\label{sec:conclusion}
In this paper we present an elaborate discussion on the capability of DUNE experiment to test the consequences of $ \mu - \tau $ reflection  symmetry considering two different modes namely, (i) 7-years of neutrino run  and (ii) 3.5-years each of neutrino and antineutrino run.  In addition, to realize $ \mu - \tau $ reflection  symmetry in the low-energy neutrino mass matrix under minimal type-I seesaw formalism, we add two heavy right-handed neutrino fields in the SM. This symmetry predicts  maximal atmospheric mixing angle (i.e., $ \theta_{23} = 45^\circ$) and Dirac CP phase  (i.e., $ \delta = \pm 90 ^\circ$) along with trivial Majorana phases in the leptonic sector. In this framework, we also find remaining oscillation parameters both analytically as well as numerically. Later, considering numerical best-fit values of neutrino oscillation parameters as our true benchmark point, we find the allowed area  in the ($ \delta - \sin^{2}\theta_{23}$) plane for DUNE. 
Further, as the latest global best-fit data prefer non-maximal $ \delta $ as well as $ \theta_{23} $, we perform our study considering  global best-fit values as one of our true benchmark point in the context of DUNE. Subsequently, we extend our study to break $ \mu - \tau $ reflection  symmetry by introducing explicit breaking term in the high-energy Dirac and Majorana neutrino mass matrices, respectively. Given the breaking scenario, we calculate the set of neutrino oscillation parameters and considering this set as the true benchmark point we find the allowed area in the test ($ \delta - \sin^{2}\theta_{23}$) plane for DUNE.
It is noteworthy to make a note here that  allowed parameter space in the test ($ \delta - \sin^{2}\theta_{23}$) plane also gives an idea about the precision of these two poorly determined parameters for DUNE.  Later, we examine the potential of DUNE to rule out maximal 
CP-violation (CPV) or CP-conservation hypothesis in each broken scenario.

We summarize DUNE's capability to test  interesting hypotheses  for all considered cases in Table~\ref{tab:Summarize}.  Given the framework of  $ \mu - \tau $ reflection symmetry, we notice that DUNE can rule out CP-conservation hypothesis at 3$ \sigma $ confidence level even with only the neutrino mode run for both the mass orderings, respectively, whereas the  DUNE[$ 3.5\nu + 3.5\overline{\nu} $] mode can reject the same at 5$ \sigma $ only in the case of IMO. Further,  considering global best-fit values as one of our cases, we find that both the considered modes of DUNE can exclude  both hypotheses at 1$ \sigma $ C.L. only for NMO, whereas it can exclude the CP-conservation hypothesis  at  5$ \sigma $ C.L. for IMO with ($ 3.5\nu + 3.5\overline{\nu} $) mode of DUNE but not in the case of NMO. Later, by inspecting  broken scenario \textbf{BS1}, we notice that DUNE can exclude the phenomenon of maximal CPV at 3$ \sigma $ C.L  but not the phenomenon of CP-conservation even at  1$ \sigma $ C.L. for NMO. Subsequently for IMO, we find that  it can rule out CP-conservation hypothesis even at 5$ \sigma $ C.L. with  DUNE[$ 3.5\nu + 3.5\overline{\nu} $] but not maximal CPV hypothesis. Moving to the \textbf{BS2} scenario, we observe that both the specifications of DUNE can exclude CP-conservation hypothesis at 3$ \sigma $ C.L. for NMO as well as IMO. Besides this, it can rule out theory of CP-conservation even at 5$ \sigma $ C.L. only for inverted mass ordering. Examining both the \textbf{BS3}, \textbf{BS4} scenarios, we come to the conclusion that DUNE can exclude either the maximal CP-violation or CP-conservation hypothesis at 1$ \sigma $ C.L. for IMO, whereas, both of the scenarios can rule out the CP-conservation hypothesis at 3$ \sigma $ C.L. only for NMO. In the case of IMO, \textbf{BS3} can rule out CP-conservation hypothesis at 3$ \sigma $ C.L., whereas \textbf{BS4} can exclude the maximal CPV hypothesis at 3$ \sigma $ C.L. considering DUNE[$ 3.5\nu + 3.5\overline{\nu} $]. In addition, by inspecting all the scenarios for both the mass orderings, we notice that none of the scenarios of NMO can exclude any of the concerned hypotheses at 5$ \sigma $ C.L. However, except for the  \textbf{BS3} and \textbf{BS4}, the remaining scenarios of IMO can exclude the CP-conservation hypothesis with DUNE[$ 3.5\nu + 3.5\overline{\nu} $] at the same confidence level.

 Afterwards, we also examine the precision of both the less-known parameters, $ \delta $, $ \theta_{23} $, and as a case study we present our results at 3$ \sigma $ confidence level in Table \ref{tab:Precision} . By scrutinizing all the possibilities, we notice that the \textbf{BS4} gives the worst precision on the Dirac CP-phase, $ \delta $ of 41.1\%  in the case of DUNE[$ 7\nu + 0\overline{\nu} $]  for NMO, whereas \textbf{BS1} comes with the best precision of 25.0\% among all concerned cases considering DUNE[$ 3.5\nu + 3.5\overline{\nu} $] for NMO. Similarly, for the 2-3 mixing angle, $ \theta_{23} $, we find that global best-fit value with DUNE[$ 7\nu + 0\overline{\nu} $] mode gives a worst precision of 11.6\% for IMO, whereas \textbf{BS2} for IMO gives a best precision of 8.5\% for DUNE[$ 3.5\nu + 3.5\overline{\nu} $]. Also, by investigating all scenarios, we notice that the scenario \textbf{BS3} is able to exclude the lower octant of $ \theta_{23} $ at 1$ \sigma $ C.L. for NMO and analysis of global best-fit value shows similar conclusion  in context of IMO. Note that results discussed here can be used to test DUNE's potential for the discrimination of different scenarios.

Finally, we conclude this work with a remark that with the available data in the neutrino oscillation sector, the $ \mu - \tau $ reflection  symmetry possesses as one of the finest theoretically favored approaches to study some intriguing aspects of neutrinos. On the other hand forthcoming experiment, like DUNE with its high statistics and ability to measure ($ \delta, \theta_{23}$) with high precision serves as an impeccable experiment to test numerous predictions of different models. 
\begin{acknowledgements}
Author is  indebtedly grateful to Prof. Zhi-zhong Xing for his insightful comments and careful reading of the manuscript. Author also likes to thank Prof. Srubabati Goswami, Prof. Shun Zhou, Dr. Jue Zhang, Dr. Rahul Srivastava and Mr. Guo-yuan Huang for useful discussions. The research work of NN  was supported in part by the National Natural Science Foundation of China under grant No. 11775231.
\end{acknowledgements}

\bibliographystyle{apsrev4-1}
\bibliography{NuFlav}

\begin{thebibliography}{53}%
\makeatletter
\providecommand \@ifxundefined [1]{%
 \@ifx{#1\undefined}
}%
\providecommand \@ifnum [1]{%
 \ifnum #1\expandafter \@firstoftwo
 \else \expandafter \@secondoftwo
 \fi
}%
\providecommand \@ifx [1]{%
 \ifx #1\expandafter \@firstoftwo
 \else \expandafter \@secondoftwo
 \fi
}%
\providecommand \natexlab [1]{#1}%
\providecommand \enquote  [1]{``#1''}%
\providecommand \bibnamefont  [1]{#1}%
\providecommand \bibfnamefont [1]{#1}%
\providecommand \citenamefont [1]{#1}%
\providecommand \href@noop [0]{\@secondoftwo}%
\providecommand \href [0]{\begingroup \@sanitize@url \@href}%
\providecommand \@href[1]{\@@startlink{#1}\@@href}%
\providecommand \@@href[1]{\endgroup#1\@@endlink}%
\providecommand \@sanitize@url [0]{\catcode `\\12\catcode `\$12\catcode
  `\&12\catcode `\#12\catcode `\^12\catcode `\_12\catcode `\%12\relax}%
\providecommand \@@startlink[1]{}%
\providecommand \@@endlink[0]{}%
\providecommand \url  [0]{\begingroup\@sanitize@url \@url }%
\providecommand \@url [1]{\endgroup\@href {#1}{\urlprefix }}%
\providecommand \urlprefix  [0]{URL }%
\providecommand \Eprint [0]{\href }%
\providecommand \doibase [0]{http://dx.doi.org/}%
\providecommand \selectlanguage [0]{\@gobble}%
\providecommand \bibinfo  [0]{\@secondoftwo}%
\providecommand \bibfield  [0]{\@secondoftwo}%
\providecommand \translation [1]{[#1]}%
\providecommand \BibitemOpen [0]{}%
\providecommand \bibitemStop [0]{}%
\providecommand \bibitemNoStop [0]{.\EOS\space}%
\providecommand \EOS [0]{\spacefactor3000\relax}%
\providecommand \BibitemShut  [1]{\csname bibitem#1\endcsname}%
\let\auto@bib@innerbib\@empty
\bibitem [{\citenamefont {Patrignani}\ \emph {et~al.}(2016)\citenamefont
  {Patrignani} \emph {et~al.}}]{Patrignani:2016xqp}%
  \BibitemOpen
  \bibfield  {author} {\bibinfo {author} {\bibfnamefont {C.}~\bibnamefont
  {Patrignani}} \emph {et~al.} (\bibinfo {collaboration} {Particle Data
  Group}),\ }\href {\doibase 10.1088/1674-1137/40/10/100001} {\bibfield
  {journal} {\bibinfo  {journal} {Chin. Phys.}\ }\textbf {\bibinfo {volume}
  {C40}},\ \bibinfo {pages} {100001} (\bibinfo {year} {2016})}\BibitemShut
  {NoStop}%
\bibitem [{\citenamefont {Minkowski}(1977)}]{Minkowski:1977sc}%
  \BibitemOpen
  \bibfield  {author} {\bibinfo {author} {\bibfnamefont {P.}~\bibnamefont
  {Minkowski}},\ }\href {\doibase 10.1016/0370-2693(77)90435-X} {\bibfield
  {journal} {\bibinfo  {journal} {Phys. Lett.}\ }\textbf {\bibinfo {volume}
  {67B}},\ \bibinfo {pages} {421} (\bibinfo {year} {1977})}\BibitemShut
  {NoStop}%
\bibitem [{\citenamefont {Yanagida}(1979)}]{Yanagida:1979as}%
  \BibitemOpen
  \bibfield  {author} {\bibinfo {author} {\bibfnamefont {T.}~\bibnamefont
  {Yanagida}},\ }\bibfield  {booktitle} {\emph {\bibinfo {booktitle}
  {{Proceedings: Workshop on the Unified Theories and the Baryon Number in the
  Universe: Tsukuba, Japan, February 13-14, 1979}}},\ }\href@noop {} {\bibfield
   {journal} {\bibinfo  {journal} {Conf. Proc.}\ }\textbf {\bibinfo {volume}
  {C7902131}},\ \bibinfo {pages} {95} (\bibinfo {year} {1979})}\BibitemShut
  {NoStop}%
\bibitem [{\citenamefont {Gell-Mann}\ \emph {et~al.}(1979)\citenamefont
  {Gell-Mann}, \citenamefont {Ramond},\ and\ \citenamefont
  {Slansky}}]{GellMann:1980vs}%
  \BibitemOpen
  \bibfield  {author} {\bibinfo {author} {\bibfnamefont {M.}~\bibnamefont
  {Gell-Mann}}, \bibinfo {author} {\bibfnamefont {P.}~\bibnamefont {Ramond}}, \
  and\ \bibinfo {author} {\bibfnamefont {R.}~\bibnamefont {Slansky}},\
  }\bibfield  {booktitle} {\emph {\bibinfo {booktitle} {{Supergravity Workshop
  Stony Brook, New York, September 27-28, 1979}}},\ }\href@noop {} {\bibfield
  {journal} {\bibinfo  {journal} {Conf. Proc.}\ }\textbf {\bibinfo {volume}
  {C790927}},\ \bibinfo {pages} {315} (\bibinfo {year} {1979})},\ \Eprint
  {http://arxiv.org/abs/1306.4669} {arXiv:1306.4669 [hep-th]} \BibitemShut
  {NoStop}%
\bibitem [{\citenamefont {Mohapatra}\ and\ \citenamefont
  {Senjanovic}(1980)}]{Mohapatra:1979ia}%
  \BibitemOpen
  \bibfield  {author} {\bibinfo {author} {\bibfnamefont {R.~N.}\ \bibnamefont
  {Mohapatra}}\ and\ \bibinfo {author} {\bibfnamefont {G.}~\bibnamefont
  {Senjanovic}},\ }\href {\doibase 10.1103/PhysRevLett.44.912} {\bibfield
  {journal} {\bibinfo  {journal} {Phys. Rev. Lett.}\ }\textbf {\bibinfo
  {volume} {44}},\ \bibinfo {pages} {912} (\bibinfo {year} {1980})}\BibitemShut
  {NoStop}%
\bibitem [{\citenamefont {Schechter}\ and\ \citenamefont
  {Valle}(1980)}]{Schechter:1980gr}%
  \BibitemOpen
  \bibfield  {author} {\bibinfo {author} {\bibfnamefont {J.}~\bibnamefont
  {Schechter}}\ and\ \bibinfo {author} {\bibfnamefont {J.~W.~F.}\ \bibnamefont
  {Valle}},\ }\href {\doibase 10.1103/PhysRevD.22.2227} {\bibfield  {journal}
  {\bibinfo  {journal} {Phys. Rev.}\ }\textbf {\bibinfo {volume} {D22}},\
  \bibinfo {pages} {2227} (\bibinfo {year} {1980})}\BibitemShut {NoStop}%
\bibitem [{\citenamefont {Altarelli}\ and\ \citenamefont
  {Feruglio}(2010)}]{Altarelli:2010gt}%
  \BibitemOpen
  \bibfield  {author} {\bibinfo {author} {\bibfnamefont {G.}~\bibnamefont
  {Altarelli}}\ and\ \bibinfo {author} {\bibfnamefont {F.}~\bibnamefont
  {Feruglio}},\ }\href {\doibase 10.1103/RevModPhys.82.2701} {\bibfield
  {journal} {\bibinfo  {journal} {Rev. Mod. Phys.}\ }\textbf {\bibinfo {volume}
  {82}},\ \bibinfo {pages} {2701} (\bibinfo {year} {2010})},\ \Eprint
  {http://arxiv.org/abs/1002.0211} {arXiv:1002.0211 [hep-ph]} \BibitemShut
  {NoStop}%
\bibitem [{\citenamefont {Altarelli}\ \emph {et~al.}(2013)\citenamefont
  {Altarelli}, \citenamefont {Feruglio},\ and\ \citenamefont
  {Merlo}}]{Altarelli:2012ss}%
  \BibitemOpen
  \bibfield  {author} {\bibinfo {author} {\bibfnamefont {G.}~\bibnamefont
  {Altarelli}}, \bibinfo {author} {\bibfnamefont {F.}~\bibnamefont {Feruglio}},
  \ and\ \bibinfo {author} {\bibfnamefont {L.}~\bibnamefont {Merlo}},\ }\href
  {\doibase 10.1002/prop.201200117} {\bibfield  {journal} {\bibinfo  {journal}
  {Fortsch. Phys.}\ }\textbf {\bibinfo {volume} {61}},\ \bibinfo {pages} {507}
  (\bibinfo {year} {2013})},\ \Eprint {http://arxiv.org/abs/1205.5133}
  {arXiv:1205.5133 [hep-ph]} \BibitemShut {NoStop}%
\bibitem [{\citenamefont {Smirnov}(2011)}]{Smirnov:2011jv}%
  \BibitemOpen
  \bibfield  {author} {\bibinfo {author} {\bibfnamefont {A.~{\relax Yu}.}\
  \bibnamefont {Smirnov}},\ }\bibfield  {booktitle} {\emph {\bibinfo
  {booktitle} {{Proceedings, 2nd Symposium on Prospects in the Physics of
  Discrete Symmetries (DISCRETE 2010): Rome, Italy, December 6-11, 2010}}},\
  }\href {\doibase 10.1088/1742-6596/335/1/012006} {\bibfield  {journal}
  {\bibinfo  {journal} {J. Phys. Conf. Ser.}\ }\textbf {\bibinfo {volume}
  {335}},\ \bibinfo {pages} {012006} (\bibinfo {year} {2011})},\ \Eprint
  {http://arxiv.org/abs/1103.3461} {arXiv:1103.3461 [hep-ph]} \BibitemShut
  {NoStop}%
\bibitem [{\citenamefont {Ishimori}\ \emph {et~al.}(2010)\citenamefont
  {Ishimori}, \citenamefont {Kobayashi}, \citenamefont {Ohki}, \citenamefont
  {Shimizu}, \citenamefont {Okada},\ and\ \citenamefont
  {Tanimoto}}]{Ishimori:2010au}%
  \BibitemOpen
  \bibfield  {author} {\bibinfo {author} {\bibfnamefont {H.}~\bibnamefont
  {Ishimori}}, \bibinfo {author} {\bibfnamefont {T.}~\bibnamefont {Kobayashi}},
  \bibinfo {author} {\bibfnamefont {H.}~\bibnamefont {Ohki}}, \bibinfo {author}
  {\bibfnamefont {Y.}~\bibnamefont {Shimizu}}, \bibinfo {author} {\bibfnamefont
  {H.}~\bibnamefont {Okada}}, \ and\ \bibinfo {author} {\bibfnamefont
  {M.}~\bibnamefont {Tanimoto}},\ }\href {\doibase 10.1143/PTPS.183.1}
  {\bibfield  {journal} {\bibinfo  {journal} {Prog. Theor. Phys. Suppl.}\
  }\textbf {\bibinfo {volume} {183}},\ \bibinfo {pages} {1} (\bibinfo {year}
  {2010})},\ \Eprint {http://arxiv.org/abs/1003.3552} {arXiv:1003.3552
  [hep-th]} \BibitemShut {NoStop}%
\bibitem [{\citenamefont {King}\ and\ \citenamefont
  {Luhn}(2013)}]{King:2013eh}%
  \BibitemOpen
  \bibfield  {author} {\bibinfo {author} {\bibfnamefont {S.~F.}\ \bibnamefont
  {King}}\ and\ \bibinfo {author} {\bibfnamefont {C.}~\bibnamefont {Luhn}},\
  }\href {\doibase 10.1088/0034-4885/76/5/056201} {\bibfield  {journal}
  {\bibinfo  {journal} {Rept. Prog. Phys.}\ }\textbf {\bibinfo {volume} {76}},\
  \bibinfo {pages} {056201} (\bibinfo {year} {2013})},\ \Eprint
  {http://arxiv.org/abs/1301.1340} {arXiv:1301.1340 [hep-ph]} \BibitemShut
  {NoStop}%
\bibitem [{\citenamefont {Harrison}\ and\ \citenamefont
  {Scott}(2002)}]{Harrison:2002et}%
  \BibitemOpen
  \bibfield  {author} {\bibinfo {author} {\bibfnamefont {P.~F.}\ \bibnamefont
  {Harrison}}\ and\ \bibinfo {author} {\bibfnamefont {W.~G.}\ \bibnamefont
  {Scott}},\ }\href {\doibase 10.1016/S0370-2693(02)02772-7} {\bibfield
  {journal} {\bibinfo  {journal} {Phys. Lett.}\ }\textbf {\bibinfo {volume}
  {B547}},\ \bibinfo {pages} {219} (\bibinfo {year} {2002})},\ \Eprint
  {http://arxiv.org/abs/hep-ph/0210197} {arXiv:hep-ph/0210197 [hep-ph]}
  \BibitemShut {NoStop}%
\bibitem [{\citenamefont {Xing}\ and\ \citenamefont
  {Zhao}(2016)}]{Xing:2015fdg}%
  \BibitemOpen
  \bibfield  {author} {\bibinfo {author} {\bibfnamefont {Z.-z.}\ \bibnamefont
  {Xing}}\ and\ \bibinfo {author} {\bibfnamefont {Z.-h.}\ \bibnamefont
  {Zhao}},\ }\href {\doibase 10.1088/0034-4885/79/7/076201} {\bibfield
  {journal} {\bibinfo  {journal} {Rept. Prog. Phys.}\ }\textbf {\bibinfo
  {volume} {79}},\ \bibinfo {pages} {076201} (\bibinfo {year} {2016})},\
  \Eprint {http://arxiv.org/abs/1512.04207} {arXiv:1512.04207 [hep-ph]}
  \BibitemShut {NoStop}%
\bibitem [{\citenamefont {Ferreira}\ \emph {et~al.}(2012)\citenamefont
  {Ferreira}, \citenamefont {Grimus}, \citenamefont {Lavoura},\ and\
  \citenamefont {Ludl}}]{Ferreira:2012ri}%
  \BibitemOpen
  \bibfield  {author} {\bibinfo {author} {\bibfnamefont {P.~M.}\ \bibnamefont
  {Ferreira}}, \bibinfo {author} {\bibfnamefont {W.}~\bibnamefont {Grimus}},
  \bibinfo {author} {\bibfnamefont {L.}~\bibnamefont {Lavoura}}, \ and\
  \bibinfo {author} {\bibfnamefont {P.~O.}\ \bibnamefont {Ludl}},\ }\href
  {\doibase 10.1007/JHEP09(2012)128} {\bibfield  {journal} {\bibinfo  {journal}
  {JHEP}\ }\textbf {\bibinfo {volume} {09}},\ \bibinfo {pages} {128} (\bibinfo
  {year} {2012})},\ \Eprint {http://arxiv.org/abs/1206.7072} {arXiv:1206.7072
  [hep-ph]} \BibitemShut {NoStop}%
\bibitem [{\citenamefont {Mohapatra}\ and\ \citenamefont
  {Nishi}(2012)}]{Mohapatra:2012tb}%
  \BibitemOpen
  \bibfield  {author} {\bibinfo {author} {\bibfnamefont {R.~N.}\ \bibnamefont
  {Mohapatra}}\ and\ \bibinfo {author} {\bibfnamefont {C.~C.}\ \bibnamefont
  {Nishi}},\ }\href {\doibase 10.1103/PhysRevD.86.073007} {\bibfield  {journal}
  {\bibinfo  {journal} {Phys. Rev.}\ }\textbf {\bibinfo {volume} {D86}},\
  \bibinfo {pages} {073007} (\bibinfo {year} {2012})},\ \Eprint
  {http://arxiv.org/abs/1208.2875} {arXiv:1208.2875 [hep-ph]} \BibitemShut
  {NoStop}%
\bibitem [{\citenamefont {Ma}\ \emph {et~al.}(2015)\citenamefont {Ma},
  \citenamefont {Natale},\ and\ \citenamefont {Popov}}]{Ma:2015gka}%
  \BibitemOpen
  \bibfield  {author} {\bibinfo {author} {\bibfnamefont {E.}~\bibnamefont
  {Ma}}, \bibinfo {author} {\bibfnamefont {A.}~\bibnamefont {Natale}}, \ and\
  \bibinfo {author} {\bibfnamefont {O.}~\bibnamefont {Popov}},\ }\href
  {\doibase 10.1016/j.physletb.2015.04.064} {\bibfield  {journal} {\bibinfo
  {journal} {Phys. Lett.}\ }\textbf {\bibinfo {volume} {B746}},\ \bibinfo
  {pages} {114} (\bibinfo {year} {2015})},\ \Eprint
  {http://arxiv.org/abs/1502.08023} {arXiv:1502.08023 [hep-ph]} \BibitemShut
  {NoStop}%
\bibitem [{\citenamefont {Ma}(2016)}]{Ma:2015fpa}%
  \BibitemOpen
  \bibfield  {author} {\bibinfo {author} {\bibfnamefont {E.}~\bibnamefont
  {Ma}},\ }\href {\doibase 10.1016/j.physletb.2015.11.049} {\bibfield
  {journal} {\bibinfo  {journal} {Phys. Lett.}\ }\textbf {\bibinfo {volume}
  {B752}},\ \bibinfo {pages} {198} (\bibinfo {year} {2016})},\ \Eprint
  {http://arxiv.org/abs/1510.02501} {arXiv:1510.02501 [hep-ph]} \BibitemShut
  {NoStop}%
\bibitem [{\citenamefont {He}\ \emph {et~al.}(2015)\citenamefont {He},
  \citenamefont {Rodejohann},\ and\ \citenamefont {Xu}}]{He:2015xha}%
  \BibitemOpen
  \bibfield  {author} {\bibinfo {author} {\bibfnamefont {H.-J.}\ \bibnamefont
  {He}}, \bibinfo {author} {\bibfnamefont {W.}~\bibnamefont {Rodejohann}}, \
  and\ \bibinfo {author} {\bibfnamefont {X.-J.}\ \bibnamefont {Xu}},\ }\href
  {\doibase 10.1016/j.physletb.2015.10.066} {\bibfield  {journal} {\bibinfo
  {journal} {Phys. Lett.}\ }\textbf {\bibinfo {volume} {B751}},\ \bibinfo
  {pages} {586} (\bibinfo {year} {2015})},\ \Eprint
  {http://arxiv.org/abs/1507.03541} {arXiv:1507.03541 [hep-ph]} \BibitemShut
  {NoStop}%
\bibitem [{\citenamefont {Joshipura}\ and\ \citenamefont
  {Patel}(2015)}]{Joshipura:2015dsa}%
  \BibitemOpen
  \bibfield  {author} {\bibinfo {author} {\bibfnamefont {A.~S.}\ \bibnamefont
  {Joshipura}}\ and\ \bibinfo {author} {\bibfnamefont {K.~M.}\ \bibnamefont
  {Patel}},\ }\href {\doibase 10.1016/j.physletb.2015.07.062} {\bibfield
  {journal} {\bibinfo  {journal} {Phys. Lett.}\ }\textbf {\bibinfo {volume}
  {B749}},\ \bibinfo {pages} {159} (\bibinfo {year} {2015})},\ \Eprint
  {http://arxiv.org/abs/1507.01235} {arXiv:1507.01235 [hep-ph]} \BibitemShut
  {NoStop}%
\bibitem [{\citenamefont {Joshipura}(2015)}]{Joshipura:2015zla}%
  \BibitemOpen
  \bibfield  {author} {\bibinfo {author} {\bibfnamefont {A.~S.}\ \bibnamefont
  {Joshipura}},\ }\href {\doibase 10.1007/JHEP11(2015)186} {\bibfield
  {journal} {\bibinfo  {journal} {JHEP}\ }\textbf {\bibinfo {volume} {11}},\
  \bibinfo {pages} {186} (\bibinfo {year} {2015})},\ \Eprint
  {http://arxiv.org/abs/1506.00455} {arXiv:1506.00455 [hep-ph]} \BibitemShut
  {NoStop}%
\bibitem [{\citenamefont {Joshipura}\ and\ \citenamefont
  {Nath}(2016)}]{Joshipura:2016hvn}%
  \BibitemOpen
  \bibfield  {author} {\bibinfo {author} {\bibfnamefont {A.~S.}\ \bibnamefont
  {Joshipura}}\ and\ \bibinfo {author} {\bibfnamefont {N.}~\bibnamefont
  {Nath}},\ }\href {\doibase 10.1103/PhysRevD.94.036008} {\bibfield  {journal}
  {\bibinfo  {journal} {Phys. Rev.}\ }\textbf {\bibinfo {volume} {D94}},\
  \bibinfo {pages} {036008} (\bibinfo {year} {2016})},\ \Eprint
  {http://arxiv.org/abs/1606.01697} {arXiv:1606.01697 [hep-ph]} \BibitemShut
  {NoStop}%
\bibitem [{\citenamefont {Nishi}\ and\ \citenamefont
  {Sánchez-Vega}(2017)}]{Nishi:2016wki}%
  \BibitemOpen
  \bibfield  {author} {\bibinfo {author} {\bibfnamefont {C.~C.}\ \bibnamefont
  {Nishi}}\ and\ \bibinfo {author} {\bibfnamefont {B.~L.}\ \bibnamefont
  {Sánchez-Vega}},\ }\href {\doibase 10.1007/JHEP01(2017)068} {\bibfield
  {journal} {\bibinfo  {journal} {JHEP}\ }\textbf {\bibinfo {volume} {01}},\
  \bibinfo {pages} {068} (\bibinfo {year} {2017})},\ \Eprint
  {http://arxiv.org/abs/1611.08282} {arXiv:1611.08282 [hep-ph]} \BibitemShut
  {NoStop}%
\bibitem [{\citenamefont {Zhao}(2017)}]{Zhao:2017yvw}%
  \BibitemOpen
  \bibfield  {author} {\bibinfo {author} {\bibfnamefont {Z.-h.}\ \bibnamefont
  {Zhao}},\ }\href {\doibase 10.1007/JHEP09(2017)023} {\bibfield  {journal}
  {\bibinfo  {journal} {JHEP}\ }\textbf {\bibinfo {volume} {09}},\ \bibinfo
  {pages} {023} (\bibinfo {year} {2017})},\ \Eprint
  {http://arxiv.org/abs/1703.04984} {arXiv:1703.04984 [hep-ph]} \BibitemShut
  {NoStop}%
\bibitem [{\citenamefont {Rodejohann}\ and\ \citenamefont
  {Xu}(2017)}]{Rodejohann:2017lre}%
  \BibitemOpen
  \bibfield  {author} {\bibinfo {author} {\bibfnamefont {W.}~\bibnamefont
  {Rodejohann}}\ and\ \bibinfo {author} {\bibfnamefont {X.-J.}\ \bibnamefont
  {Xu}},\ }\href {\doibase 10.1103/PhysRevD.96.055039} {\bibfield  {journal}
  {\bibinfo  {journal} {Phys. Rev.}\ }\textbf {\bibinfo {volume} {D96}},\
  \bibinfo {pages} {055039} (\bibinfo {year} {2017})},\ \Eprint
  {http://arxiv.org/abs/1705.02027} {arXiv:1705.02027 [hep-ph]} \BibitemShut
  {NoStop}%
\bibitem [{\citenamefont {Liu}\ \emph {et~al.}(2017)\citenamefont {Liu},
  \citenamefont {Yue},\ and\ \citenamefont {Zhao}}]{Liu:2017frs}%
  \BibitemOpen
  \bibfield  {author} {\bibinfo {author} {\bibfnamefont {Z.-C.}\ \bibnamefont
  {Liu}}, \bibinfo {author} {\bibfnamefont {C.-X.}\ \bibnamefont {Yue}}, \ and\
  \bibinfo {author} {\bibfnamefont {Z.-h.}\ \bibnamefont {Zhao}},\ }\href
  {\doibase 10.1007/JHEP10(2017)102} {\bibfield  {journal} {\bibinfo  {journal}
  {JHEP}\ }\textbf {\bibinfo {volume} {10}},\ \bibinfo {pages} {102} (\bibinfo
  {year} {2017})},\ \Eprint {http://arxiv.org/abs/1707.05535} {arXiv:1707.05535
  [hep-ph]} \BibitemShut {NoStop}%
\bibitem [{\citenamefont {Xing}\ \emph {et~al.}(2017)\citenamefont {Xing},
  \citenamefont {Zhang},\ and\ \citenamefont {Zhu}}]{Xing:2017mkx}%
  \BibitemOpen
  \bibfield  {author} {\bibinfo {author} {\bibfnamefont {Z.-z.}\ \bibnamefont
  {Xing}}, \bibinfo {author} {\bibfnamefont {D.}~\bibnamefont {Zhang}}, \ and\
  \bibinfo {author} {\bibfnamefont {J.-y.}\ \bibnamefont {Zhu}},\ }\href
  {\doibase 10.1007/JHEP11(2017)135} {\bibfield  {journal} {\bibinfo  {journal}
  {JHEP}\ }\textbf {\bibinfo {volume} {11}},\ \bibinfo {pages} {135} (\bibinfo
  {year} {2017})},\ \Eprint {http://arxiv.org/abs/1708.09144} {arXiv:1708.09144
  [hep-ph]} \BibitemShut {NoStop}%
\bibitem [{\citenamefont {Xing}\ and\ \citenamefont
  {Zhu}(2017)}]{Xing:2017cwb}%
  \BibitemOpen
  \bibfield  {author} {\bibinfo {author} {\bibfnamefont {Z.-z.}\ \bibnamefont
  {Xing}}\ and\ \bibinfo {author} {\bibfnamefont {J.-y.}\ \bibnamefont {Zhu}},\
  }\href {\doibase 10.1088/1674-1137/41/12/123103} {\bibfield  {journal}
  {\bibinfo  {journal} {Chin. Phys.}\ }\textbf {\bibinfo {volume} {C41}},\
  \bibinfo {pages} {123103} (\bibinfo {year} {2017})},\ \Eprint
  {http://arxiv.org/abs/1707.03676} {arXiv:1707.03676 [hep-ph]} \BibitemShut
  {NoStop}%
\bibitem [{\citenamefont {Nath}\ \emph {et~al.}(2018)\citenamefont {Nath},
  \citenamefont {Xing},\ and\ \citenamefont {Zhang}}]{Nath:2018hjx}%
  \BibitemOpen
  \bibfield  {author} {\bibinfo {author} {\bibfnamefont {N.}~\bibnamefont
  {Nath}}, \bibinfo {author} {\bibfnamefont {Z.-z.}\ \bibnamefont {Xing}}, \
  and\ \bibinfo {author} {\bibfnamefont {J.}~\bibnamefont {Zhang}},\ }\href
  {\doibase 10.1140/epjc/s10052-018-5751-y} {\bibfield  {journal} {\bibinfo
  {journal} {Eur. Phys. J.}\ }\textbf {\bibinfo {volume} {C78}},\ \bibinfo
  {pages} {289} (\bibinfo {year} {2018})},\ \Eprint
  {http://arxiv.org/abs/1801.09931} {arXiv:1801.09931 [hep-ph]} \BibitemShut
  {NoStop}%
\bibitem [{\citenamefont {Zhao}(2018)}]{Zhao:2018vxy}%
  \BibitemOpen
  \bibfield  {author} {\bibinfo {author} {\bibfnamefont {Z.-h.}\ \bibnamefont
  {Zhao}},\ }\href@noop {} {\  (\bibinfo {year} {2018})},\ \Eprint
  {http://arxiv.org/abs/1803.04603} {arXiv:1803.04603 [hep-ph]} \BibitemShut
  {NoStop}%
\bibitem [{\citenamefont {Chakraborty}\ \emph {et~al.}(2018)\citenamefont
  {Chakraborty}, \citenamefont {Deepthi}, \citenamefont {Goswami},
  \citenamefont {Joshipura},\ and\ \citenamefont {Nath}}]{Chakraborty:2018dew}%
  \BibitemOpen
  \bibfield  {author} {\bibinfo {author} {\bibfnamefont {K.}~\bibnamefont
  {Chakraborty}}, \bibinfo {author} {\bibfnamefont {K.~N.}\ \bibnamefont
  {Deepthi}}, \bibinfo {author} {\bibfnamefont {S.}~\bibnamefont {Goswami}},
  \bibinfo {author} {\bibfnamefont {A.~S.}\ \bibnamefont {Joshipura}}, \ and\
  \bibinfo {author} {\bibfnamefont {N.}~\bibnamefont {Nath}},\ }\href@noop {}
  {\  (\bibinfo {year} {2018})},\ \Eprint {http://arxiv.org/abs/1804.02022}
  {arXiv:1804.02022 [hep-ph]} \BibitemShut {NoStop}%
\bibitem [{\citenamefont {Nath}(2018)}]{Nath:2018xih}%
  \BibitemOpen
  \bibfield  {author} {\bibinfo {author} {\bibfnamefont {N.}~\bibnamefont
  {Nath}},\ }\href@noop {} {\  (\bibinfo {year} {2018})},\ \Eprint
  {http://arxiv.org/abs/1808.05062} {arXiv:1808.05062 [hep-ph]} \BibitemShut
  {NoStop}%
\bibitem [{\citenamefont {Guo}\ \emph {et~al.}(2007)\citenamefont {Guo},
  \citenamefont {Xing},\ and\ \citenamefont {Zhou}}]{Guo:2006qa}%
  \BibitemOpen
  \bibfield  {author} {\bibinfo {author} {\bibfnamefont {W.-l.}\ \bibnamefont
  {Guo}}, \bibinfo {author} {\bibfnamefont {Z.-z.}\ \bibnamefont {Xing}}, \
  and\ \bibinfo {author} {\bibfnamefont {S.}~\bibnamefont {Zhou}},\ }\href
  {\doibase 10.1142/S0218301307004898} {\bibfield  {journal} {\bibinfo
  {journal} {Int. J. Mod. Phys.}\ }\textbf {\bibinfo {volume} {E16}},\ \bibinfo
  {pages} {1} (\bibinfo {year} {2007})},\ \Eprint
  {http://arxiv.org/abs/hep-ph/0612033} {arXiv:hep-ph/0612033 [hep-ph]}
  \BibitemShut {NoStop}%
\bibitem [{\citenamefont {Acciarri}\ \emph {et~al.}(2015)\citenamefont
  {Acciarri} \emph {et~al.}}]{Acciarri:2015uup}%
  \BibitemOpen
  \bibfield  {author} {\bibinfo {author} {\bibfnamefont {R.}~\bibnamefont
  {Acciarri}} \emph {et~al.} (\bibinfo {collaboration} {DUNE}),\ }\href@noop {}
  {\  (\bibinfo {year} {2015})},\ \Eprint {http://arxiv.org/abs/1512.06148}
  {arXiv:1512.06148 [physics.ins-det]} \BibitemShut {NoStop}%
\bibitem [{nuf()}]{nufit18}%
  \BibitemOpen
  \href@noop {} {\ }\bibinfo {note} {NuFIT 3.2 collaboration (2018),
  \url{http://www.nu-fit.org/}}\BibitemShut {NoStop}%
\bibitem [{\citenamefont {Esteban}\ \emph {et~al.}(2017)\citenamefont
  {Esteban}, \citenamefont {Gonzalez-Garcia}, \citenamefont {Maltoni},
  \citenamefont {Martinez-Soler},\ and\ \citenamefont
  {Schwetz}}]{Esteban:2016qun}%
  \BibitemOpen
  \bibfield  {author} {\bibinfo {author} {\bibfnamefont {I.}~\bibnamefont
  {Esteban}}, \bibinfo {author} {\bibfnamefont {M.~C.}\ \bibnamefont
  {Gonzalez-Garcia}}, \bibinfo {author} {\bibfnamefont {M.}~\bibnamefont
  {Maltoni}}, \bibinfo {author} {\bibfnamefont {I.}~\bibnamefont
  {Martinez-Soler}}, \ and\ \bibinfo {author} {\bibfnamefont {T.}~\bibnamefont
  {Schwetz}},\ }\href {\doibase 10.1007/JHEP01(2017)087} {\bibfield  {journal}
  {\bibinfo  {journal} {JHEP}\ }\textbf {\bibinfo {volume} {01}},\ \bibinfo
  {pages} {087} (\bibinfo {year} {2017})},\ \Eprint
  {http://arxiv.org/abs/1611.01514} {arXiv:1611.01514 [hep-ph]} \BibitemShut
  {NoStop}%
\bibitem [{\citenamefont {Abe}\ \emph {et~al.}(2017)\citenamefont {Abe} \emph
  {et~al.}}]{Abe:2017uxa}%
  \BibitemOpen
  \bibfield  {author} {\bibinfo {author} {\bibfnamefont {K.}~\bibnamefont
  {Abe}} \emph {et~al.} (\bibinfo {collaboration} {T2K}),\ }\href {\doibase
  10.1103/PhysRevLett.118.151801} {\bibfield  {journal} {\bibinfo  {journal}
  {Phys. Rev. Lett.}\ }\textbf {\bibinfo {volume} {118}},\ \bibinfo {pages}
  {151801} (\bibinfo {year} {2017})},\ \Eprint
  {http://arxiv.org/abs/1701.00432} {arXiv:1701.00432 [hep-ex]} \BibitemShut
  {NoStop}%
\bibitem [{\citenamefont {Radovic}\ \emph {et~al.}(2018)\citenamefont {Radovic}
  \emph {et~al.}}]{NOVA2018}%
  \BibitemOpen
  \bibfield  {author} {\bibinfo {author} {\bibfnamefont {A.}~\bibnamefont
  {Radovic}} \emph {et~al.} (\bibinfo {collaboration} {NO$\nu$A})\ }(\bibinfo
  {year} {2018})\ \bibinfo {note} {talk given at the Fermilab, January 2018,
  USA,
  \url{http://nova-docdb.fnal.gov/cgi-bin/ShowDocument?docid=25938}}\BibitemShut
  {NoStop}%
\bibitem [{\citenamefont {de~Adelhart~Toorop}\ \emph
  {et~al.}(2011)\citenamefont {de~Adelhart~Toorop}, \citenamefont {Feruglio},\
  and\ \citenamefont {Hagedorn}}]{Toorop:2011jn}%
  \BibitemOpen
  \bibfield  {author} {\bibinfo {author} {\bibfnamefont {R.}~\bibnamefont
  {de~Adelhart~Toorop}}, \bibinfo {author} {\bibfnamefont {F.}~\bibnamefont
  {Feruglio}}, \ and\ \bibinfo {author} {\bibfnamefont {C.}~\bibnamefont
  {Hagedorn}},\ }\href {\doibase 10.1016/j.physletb.2011.08.013} {\bibfield
  {journal} {\bibinfo  {journal} {Phys. Lett.}\ }\textbf {\bibinfo {volume}
  {B703}},\ \bibinfo {pages} {447} (\bibinfo {year} {2011})},\ \Eprint
  {http://arxiv.org/abs/1107.3486} {arXiv:1107.3486 [hep-ph]} \BibitemShut
  {NoStop}%
\bibitem [{\citenamefont {Hanlon}\ \emph
  {et~al.}(2014{\natexlab{a}})\citenamefont {Hanlon}, \citenamefont {Ge},\ and\
  \citenamefont {Repko}}]{Hanlon:2013ska}%
  \BibitemOpen
  \bibfield  {author} {\bibinfo {author} {\bibfnamefont {A.~D.}\ \bibnamefont
  {Hanlon}}, \bibinfo {author} {\bibfnamefont {S.-F.}\ \bibnamefont {Ge}}, \
  and\ \bibinfo {author} {\bibfnamefont {W.~W.}\ \bibnamefont {Repko}},\ }\href
  {\doibase 10.1016/j.physletb.2013.12.063} {\bibfield  {journal} {\bibinfo
  {journal} {Phys. Lett.}\ }\textbf {\bibinfo {volume} {B729}},\ \bibinfo
  {pages} {185} (\bibinfo {year} {2014}{\natexlab{a}})},\ \Eprint
  {http://arxiv.org/abs/1308.6522} {arXiv:1308.6522 [hep-ph]} \BibitemShut
  {NoStop}%
\bibitem [{\citenamefont {Hanlon}\ \emph
  {et~al.}(2014{\natexlab{b}})\citenamefont {Hanlon}, \citenamefont {Repko},\
  and\ \citenamefont {Dicus}}]{Hanlon:2014bga}%
  \BibitemOpen
  \bibfield  {author} {\bibinfo {author} {\bibfnamefont {A.~D.}\ \bibnamefont
  {Hanlon}}, \bibinfo {author} {\bibfnamefont {W.~W.}\ \bibnamefont {Repko}}, \
  and\ \bibinfo {author} {\bibfnamefont {D.~A.}\ \bibnamefont {Dicus}},\ }\href
  {\doibase 10.1155/2014/469572} {\bibfield  {journal} {\bibinfo  {journal}
  {Adv. High Energy Phys.}\ }\textbf {\bibinfo {volume} {2014}},\ \bibinfo
  {pages} {469572} (\bibinfo {year} {2014}{\natexlab{b}})},\ \Eprint
  {http://arxiv.org/abs/1403.7552} {arXiv:1403.7552 [hep-ph]} \BibitemShut
  {NoStop}%
\bibitem [{\citenamefont {Srivastava}\ \emph
  {et~al.}(2018{\natexlab{a}})\citenamefont {Srivastava}, \citenamefont
  {Ternes}, \citenamefont {Tórtola},\ and\ \citenamefont
  {Valle}}]{Srivastava:2017sno}%
  \BibitemOpen
  \bibfield  {author} {\bibinfo {author} {\bibfnamefont {R.}~\bibnamefont
  {Srivastava}}, \bibinfo {author} {\bibfnamefont {C.~A.}\ \bibnamefont
  {Ternes}}, \bibinfo {author} {\bibfnamefont {M.}~\bibnamefont {Tórtola}}, \
  and\ \bibinfo {author} {\bibfnamefont {J.~W.~F.}\ \bibnamefont {Valle}},\
  }\href {\doibase 10.1016/j.physletb.2018.01.014} {\bibfield  {journal}
  {\bibinfo  {journal} {Phys. Lett.}\ }\textbf {\bibinfo {volume} {B778}},\
  \bibinfo {pages} {459} (\bibinfo {year} {2018}{\natexlab{a}})},\ \Eprint
  {http://arxiv.org/abs/1711.10318} {arXiv:1711.10318 [hep-ph]} \BibitemShut
  {NoStop}%
\bibitem [{\citenamefont {Agarwalla}\ \emph {et~al.}(2018)\citenamefont
  {Agarwalla}, \citenamefont {Chatterjee}, \citenamefont {Petcov},\ and\
  \citenamefont {Titov}}]{Agarwalla:2017wct}%
  \BibitemOpen
  \bibfield  {author} {\bibinfo {author} {\bibfnamefont {S.~K.}\ \bibnamefont
  {Agarwalla}}, \bibinfo {author} {\bibfnamefont {S.~S.}\ \bibnamefont
  {Chatterjee}}, \bibinfo {author} {\bibfnamefont {S.~T.}\ \bibnamefont
  {Petcov}}, \ and\ \bibinfo {author} {\bibfnamefont {A.~V.}\ \bibnamefont
  {Titov}},\ }\href {\doibase 10.1140/epjc/s10052-018-5772-6} {\bibfield
  {journal} {\bibinfo  {journal} {Eur. Phys. J.}\ }\textbf {\bibinfo {volume}
  {C78}},\ \bibinfo {pages} {286} (\bibinfo {year} {2018})},\ \Eprint
  {http://arxiv.org/abs/1711.02107} {arXiv:1711.02107 [hep-ph]} \BibitemShut
  {NoStop}%
\bibitem [{\citenamefont {Chatterjee}\ \emph {et~al.}(2017)\citenamefont
  {Chatterjee}, \citenamefont {Masud}, \citenamefont {Pasquini},\ and\
  \citenamefont {Valle}}]{Chatterjee:2017ilf}%
  \BibitemOpen
  \bibfield  {author} {\bibinfo {author} {\bibfnamefont {S.~S.}\ \bibnamefont
  {Chatterjee}}, \bibinfo {author} {\bibfnamefont {M.}~\bibnamefont {Masud}},
  \bibinfo {author} {\bibfnamefont {P.}~\bibnamefont {Pasquini}}, \ and\
  \bibinfo {author} {\bibfnamefont {J.~W.~F.}\ \bibnamefont {Valle}},\ }\href
  {\doibase 10.1016/j.physletb.2017.09.052} {\bibfield  {journal} {\bibinfo
  {journal} {Phys. Lett.}\ }\textbf {\bibinfo {volume} {B774}},\ \bibinfo
  {pages} {179} (\bibinfo {year} {2017})},\ \Eprint
  {http://arxiv.org/abs/1708.03290} {arXiv:1708.03290 [hep-ph]} \BibitemShut
  {NoStop}%
\bibitem [{\citenamefont {Pasquini}(2018)}]{Pasquini:2018udd}%
  \BibitemOpen
  \bibfield  {author} {\bibinfo {author} {\bibfnamefont {P.}~\bibnamefont
  {Pasquini}},\ }\href@noop {} {\  (\bibinfo {year} {2018})},\ \Eprint
  {http://arxiv.org/abs/1802.00821} {arXiv:1802.00821 [hep-ph]} \BibitemShut
  {NoStop}%
\bibitem [{\citenamefont {Srivastava}\ \emph
  {et~al.}(2018{\natexlab{b}})\citenamefont {Srivastava}, \citenamefont
  {Ternes}, \citenamefont {Tórtola},\ and\ \citenamefont
  {Valle}}]{Srivastava:2018ser}%
  \BibitemOpen
  \bibfield  {author} {\bibinfo {author} {\bibfnamefont {R.}~\bibnamefont
  {Srivastava}}, \bibinfo {author} {\bibfnamefont {C.~A.}\ \bibnamefont
  {Ternes}}, \bibinfo {author} {\bibfnamefont {M.}~\bibnamefont {Tórtola}}, \
  and\ \bibinfo {author} {\bibfnamefont {J.~W.~F.}\ \bibnamefont {Valle}},\
  }\href@noop {} {\  (\bibinfo {year} {2018}{\natexlab{b}})},\ \Eprint
  {http://arxiv.org/abs/1803.10247} {arXiv:1803.10247 [hep-ph]} \BibitemShut
  {NoStop}%
\bibitem [{\citenamefont {Petcov}\ and\ \citenamefont
  {Titov}(2018)}]{Petcov:2018snn}%
  \BibitemOpen
  \bibfield  {author} {\bibinfo {author} {\bibfnamefont {S.~T.}\ \bibnamefont
  {Petcov}}\ and\ \bibinfo {author} {\bibfnamefont {A.~V.}\ \bibnamefont
  {Titov}},\ }\href@noop {} {\  (\bibinfo {year} {2018})},\ \Eprint
  {http://arxiv.org/abs/1804.00182} {arXiv:1804.00182 [hep-ph]} \BibitemShut
  {NoStop}%
\bibitem [{\citenamefont {Feroz}\ and\ \citenamefont
  {Hobson}(2008)}]{Feroz:2007kg}%
  \BibitemOpen
  \bibfield  {author} {\bibinfo {author} {\bibfnamefont {F.}~\bibnamefont
  {Feroz}}\ and\ \bibinfo {author} {\bibfnamefont {M.~P.}\ \bibnamefont
  {Hobson}},\ }\href {\doibase 10.1111/j.1365-2966.2007.12353.x} {\bibfield
  {journal} {\bibinfo  {journal} {Mon. Not. Roy. Astron. Soc.}\ }\textbf
  {\bibinfo {volume} {384}},\ \bibinfo {pages} {449} (\bibinfo {year}
  {2008})},\ \Eprint {http://arxiv.org/abs/0704.3704} {arXiv:0704.3704
  [astro-ph]} \BibitemShut {NoStop}%
\bibitem [{\citenamefont {Feroz}\ \emph {et~al.}(2009)\citenamefont {Feroz},
  \citenamefont {Hobson},\ and\ \citenamefont {Bridges}}]{Feroz:2008xx}%
  \BibitemOpen
  \bibfield  {author} {\bibinfo {author} {\bibfnamefont {F.}~\bibnamefont
  {Feroz}}, \bibinfo {author} {\bibfnamefont {M.~P.}\ \bibnamefont {Hobson}}, \
  and\ \bibinfo {author} {\bibfnamefont {M.}~\bibnamefont {Bridges}},\ }\href
  {\doibase 10.1111/j.1365-2966.2009.14548.x} {\bibfield  {journal} {\bibinfo
  {journal} {Mon. Not. Roy. Astron. Soc.}\ }\textbf {\bibinfo {volume} {398}},\
  \bibinfo {pages} {1601} (\bibinfo {year} {2009})},\ \Eprint
  {http://arxiv.org/abs/0809.3437} {arXiv:0809.3437 [astro-ph]} \BibitemShut
  {NoStop}%
\bibitem [{\citenamefont {Feroz}\ \emph {et~al.}(2013)\citenamefont {Feroz},
  \citenamefont {Hobson}, \citenamefont {Cameron},\ and\ \citenamefont
  {Pettitt}}]{Feroz:2013hea}%
  \BibitemOpen
  \bibfield  {author} {\bibinfo {author} {\bibfnamefont {F.}~\bibnamefont
  {Feroz}}, \bibinfo {author} {\bibfnamefont {M.~P.}\ \bibnamefont {Hobson}},
  \bibinfo {author} {\bibfnamefont {E.}~\bibnamefont {Cameron}}, \ and\
  \bibinfo {author} {\bibfnamefont {A.~N.}\ \bibnamefont {Pettitt}},\
  }\href@noop {} {\  (\bibinfo {year} {2013})},\ \Eprint
  {http://arxiv.org/abs/1306.2144} {arXiv:1306.2144 [astro-ph.IM]} \BibitemShut
  {NoStop}%
\bibitem [{\citenamefont {Alion}\ \emph {et~al.}(2016)\citenamefont {Alion}
  \emph {et~al.}}]{Alion:2016uaj}%
  \BibitemOpen
  \bibfield  {author} {\bibinfo {author} {\bibfnamefont {T.}~\bibnamefont
  {Alion}} \emph {et~al.} (\bibinfo {collaboration} {DUNE}),\ }\href@noop {} {\
   (\bibinfo {year} {2016})},\ \Eprint {http://arxiv.org/abs/1606.09550}
  {arXiv:1606.09550 [physics.ins-det]} \BibitemShut {NoStop}%
\bibitem [{\citenamefont {Huber}\ \emph {et~al.}(2005)\citenamefont {Huber},
  \citenamefont {Lindner},\ and\ \citenamefont {Winter}}]{globes1}%
  \BibitemOpen
  \bibfield  {author} {\bibinfo {author} {\bibfnamefont {P.}~\bibnamefont
  {Huber}}, \bibinfo {author} {\bibfnamefont {M.}~\bibnamefont {Lindner}}, \
  and\ \bibinfo {author} {\bibfnamefont {W.}~\bibnamefont {Winter}},\ }\href
  {\doibase 10.1016/j.cpc.2005.01.003} {\bibfield  {journal} {\bibinfo
  {journal} {Comput. Phys. Commun.}\ }\textbf {\bibinfo {volume} {167}},\
  \bibinfo {pages} {195} (\bibinfo {year} {2005})},\ \Eprint
  {http://arxiv.org/abs/hep-ph/0407333} {arXiv:hep-ph/0407333} \BibitemShut
  {NoStop}%
\bibitem [{\citenamefont {Huber}\ \emph {et~al.}(2007)\citenamefont {Huber},
  \citenamefont {Kopp}, \citenamefont {Lindner}, \citenamefont {Rolinec},\ and\
  \citenamefont {Winter}}]{globes2}%
  \BibitemOpen
  \bibfield  {author} {\bibinfo {author} {\bibfnamefont {P.}~\bibnamefont
  {Huber}}, \bibinfo {author} {\bibfnamefont {J.}~\bibnamefont {Kopp}},
  \bibinfo {author} {\bibfnamefont {M.}~\bibnamefont {Lindner}}, \bibinfo
  {author} {\bibfnamefont {M.}~\bibnamefont {Rolinec}}, \ and\ \bibinfo
  {author} {\bibfnamefont {W.}~\bibnamefont {Winter}},\ }\href {\doibase
  10.1016/j.cpc.2007.05.004} {\bibfield  {journal} {\bibinfo  {journal}
  {Comput. Phys. Commun.}\ }\textbf {\bibinfo {volume} {177}},\ \bibinfo
  {pages} {432} (\bibinfo {year} {2007})},\ \Eprint
  {http://arxiv.org/abs/hep-ph/0701187} {arXiv:hep-ph/0701187} \BibitemShut
  {NoStop}%
\bibitem [{\citenamefont {Nath}\ \emph {et~al.}(2016)\citenamefont {Nath},
  \citenamefont {Ghosh},\ and\ \citenamefont {Goswami}}]{Nath:2015kjg}%
  \BibitemOpen
  \bibfield  {author} {\bibinfo {author} {\bibfnamefont {N.}~\bibnamefont
  {Nath}}, \bibinfo {author} {\bibfnamefont {M.}~\bibnamefont {Ghosh}}, \ and\
  \bibinfo {author} {\bibfnamefont {S.}~\bibnamefont {Goswami}},\ }\href
  {\doibase 10.1016/j.nuclphysb.2016.09.017} {\bibfield  {journal} {\bibinfo
  {journal} {Nucl. Phys.}\ }\textbf {\bibinfo {volume} {B913}},\ \bibinfo
  {pages} {381} (\bibinfo {year} {2016})},\ \Eprint
  {http://arxiv.org/abs/1511.07496} {arXiv:1511.07496 [hep-ph]} \BibitemShut
  {NoStop}%
\end{thebibliography}%
\end{document}